\documentclass{aa}
\usepackage{rotating}
\usepackage{graphicx}
\usepackage{txfonts}
\begin{document}

\title{A detailed X-ray investigation of $\zeta$\,Puppis\newline III. A spectral analysis of the whole RGS spectrum \thanks{Based on observations collected with {\it XMM-Newton}, an ESA Science Mission with instruments and contributions directly funded by ESA Member States and the USA (NASA).}}
\author{A.\ Herv\'e\inst{1} \and G.\ Rauw\inst{1} \and Y.\ Naz\'e\inst{1}\fnmsep\thanks{Research Associate FRS-FNRS (Belgium)}}
\offprints{A.\ Herv\'e}
\mail{herve@astro.ulg.ac.be}
\institute{Groupe d'Astrophysique des Hautes Energies, Institut d'Astrophysique et de G\'eophysique, Universit\'e de Li\`ege, All\'ee du 6 Ao\^ut, B\^at B5c, 4000 Li\`ege, Belgium}
\date{}
\abstract{$\zeta$\,Pup is the X-ray brightest O-type star of the sky. This object was regularly observed with the RGS instrument aboard XMM-Newton for calibration purposes, leading to an unprecedented set of high-quality spectra.}{We have previously reduced and extracted this data set and combined it into the most detailed high-resolution X-ray spectrum of any early-type star so far. Here we present the analysis of this spectrum accounting for the presence of structures in the stellar wind.}{For this purpose, we use our new modeling tool that allows fitting the entire spectrum with a multi-temperature plasma. We illustrate the impact of a proper treatment of the radial dependence of the X-ray opacity of the cool wind on the best-fit radial distribution of the temperature of the X-ray plasma.}{The best fit of the RGS spectrum of $\zeta$\,Pup is obtained assuming no porosity. Four plasma components at temperatures between 0.10 and 0.69\,keV are needed to adequately represent the observed spectrum. Whilst the hardest emission is concentrated between $\sim 3$ and 4\,R$_*$, the softer emission starts already at 1.5\,R$_*$ and extends to the outer regions of the wind.}{The inferred radial distribution of the plasma temperatures agrees rather well with theoretical expectations. The mass-loss rate and CNO abundances corresponding to our best-fit model also agree quite well with the results of recent studies of $\zeta$\,Pup in the UV and optical domain.}
\authorrunning{Herv\'e et al.}
\titlerunning{A detailed X-ray investigation of $\zeta$-Puppis}
\maketitle

\section{Introduction}
Through their powerful stellar winds, and their violent deaths as supernovae, massive stars have a tremendous impact on the ecology of their galaxies. The feedback by the stellar winds of early-type stars is indeed a major ingredient of the chemical evolution of the interstellar medium. Moreover, the mass-loss has an important impact on the evolution of the star. The quantity of material released by the wind as well as its chemical composition depend on the evolutionary stage of the star. A proper understanding of the stellar winds of massive stars is therefore fundamental for our knowledge of the stellar evolution and the impact on the interstellar medium. However, the determination of the mass-loss rate of a star is a complicated task. 

Over the last decade, the importance of the structures of the stellar winds and their impact on the determination of the mass-loss rates became more and more obvious. Observationally, these structures manifest themselves through low-amplitude variability of emission lines such as H$\alpha$ (Eversberg et al.\ \cite{ELM}), and through difficulties to fit the observed spectra with model atmosphere codes that assume homogeneous winds (Bouret et al.\ \cite{BLH}). These small-scale structures are usually attributed to hydrodynamic instabilities intrinsic to the winds (Feldmeier et al. \cite{feld97}).

Accounting for the fragmented structure of the winds leads to a downward revision of the mass-loss rates by a factor between a few and ten. Owing to the different sensitivities of the spectra over various wavelength ranges to the structures, multi-wavelength analyses are required to determine consistent mass-loss rates of massive stars. For instance, H$\alpha$ emission and free-free radio emission are collisional processes which scale with the wind-density squared and are thus sensitive to the clumping factor regardless of the size of the clumps. In clumped winds with a void inter-clump medium, the wind's velocity law misses some values along a given sight line. This porosity in the velocity field is called `vorosity' (Owocki \cite{owockiliege}). The latter modifies the Sobolev length and its impact is mainly visible on the UV resonance lines. In a fragmented wind with non-optically thin clumps, X-ray lines are sensitive to porosity, which depends on clump continuum optical depth, and thus clump size as well as the clumping factor. 

The X-ray line profiles can also give information on the hydrodynamical instabilities in the wind. In fact, considering the wind embedded shock scenario, the emission of X-ray photons arises when two fragments with different velocities collide (Lucy \& White \cite{lucy80}). The observable morphology of an X-ray line depends on the interplay of emission by the hot plasma and photoelectric absorption by the cool wind component. The latter effect is obviously also sensitive to the presence or absence of structures in the wind. On the other hand, the location of the X-ray emission region as well as the temperature of the X-ray emitting plasma give important information on the hydrodynamical properties of the winds.

$\zeta$\,Pup (= HD\,66811, O4\,Ief) has been the target of numerous studies over a wide range of wavelengths. This star is often considered as a cornerstone object for the understanding of massive stars. The analysis of the far UV - UV - optical spectra by Bouret et al.\ (\cite{bouret12}) confirmed a super-solar total CNO abundance, with N being highly overabundant and C being strongly depleted. On the other hand, Pauldrach et al.\ (\cite{pauldrach12}) found quite different results in their far UV - UV spectral modeling, indicating an over-abundance of both nitrogen and carbon and a strong depletion of oxygen. Also, they determine a homogeneous mass-loss rate ($\sim 13 \times 10^{-6}$\,M$_{\odot}$\,yr$^{-1}$) about six times larger than the clumped mass-loss rates determined by Bouret et al.\ (\cite{bouret12}) and Najarro et al.\ (\cite{najarro11}). This situation is embarrassing as one would aim for consistent parameters obtained from studies using different wavebands and different modeling tools. 
 
$\zeta$\,Pup is also a bright X-ray source and has been observed both with {\it Chandra} and {\it XMM-Newton} at high spectral resolution. The star was indeed regularly observed with the RGS instrument aboard {\it XMM-Newton} for calibration purposes, yielding a combined data set of unprecedented quality (Naz\'e et al.\ \cite{naze12}). Previous studies of $\zeta$\,Pup's high-resolution X-ray spectrum revealed new interesting results (Cassinelli et al.\ \cite{cassinelli01}, Kahn et al.\ \cite{kahn01}, Oskinova et al. \cite{OFH}, Leutenegger et al.\ \cite{Leutenegger06}, \cite{Leutenegger07}, Cohen et al.\ \cite{cohen10}). For instance, the analyses by Cassinelli et al.\ (\cite{cassinelli01}) and Leutenegger et al.\ (\cite{Leutenegger06}) of the $forbiden, inter-combination, resonance$ ({\it{fir}}) triplets of He-like ions revealed that the inner boundary of the X-ray emission region is rather close to the star (at about 1.5 R$_*$ and possibly even closer for the hottest plasma, Cassinelli et al.\ \cite{cassinelli01}). However, these studies were done on individual line profiles considered separately. Each specific line was in some sense assumed to be formed by a plasma at the temperature of maximum emissivity of the line. This approach thus neglects the fact that several plasma components (each one characterized by its specific temperature, emission measure and location inside the wind) may contribute significantly to a given line. To the best of our knowledge, there have been no previous attempts to fit the entire high-resolution X-ray spectrum (either RGS or HETG) of an O-type star at once by properly combining the contributions from several hot plasma components and modelling the absorption by the cool wind, accounting explicitly for the geometry of the problem.

This paper thus explores, for the first time, the possibility to fit all the lines of an X-ray spectrum between 6 and 35\,\AA\ simultaneously and coherently, assuming a discrete distribution of plasma components, characterized by their own temperature and location in the wind. To do so, we use the code developed by Herv\'e et al.\ (\cite{herve12}) which accounts for the intrinsic emissivity of the hot plasma as a function of temperature and chemical composition. We further introduce a multi-temperature plasma spectral fitting procedure. The goal is to determine the properties of the X-ray emission region accounting for the most recent results from other wavelength domains. These properties can ultimately be compared to the predictions of existing and future hydrodynamical simulations of the instabilities of stellar winds. 

This paper is organized as follows. In Sect\,\ref{Observations}, we present a short summary of the $\zeta$\,Pup  RGS dataset. Then we briefly recall the main assumptions in our modelling tool (Herv\'e et al.\ \cite{herve12}) concerning the intrinsic emissivity and the absorption formalism (Sect.\,\ref{Code}). We also present, in Sect.\,\ref{multiT}, the different physical processes we have included in the code to better reproduce the observational spectrum of $\zeta$\,Pup. Finally, in Sect.\,\ref{Results}, we discuss the results that we obtain, especially as far as the composition and the location of the different plasma components in the wind are concerned. 

\section{Observational data}\label{Observations}
Naz\'e et al.\ (\cite{naze12}) extracted and reduced 18 observations of $\zeta$\,Pup with {\it XMM-Newton} which yield more than 700\,ks of useful exposure. For the details of the data reduction, we refer the reader to the paper of Naz\'e et al.\ (\cite{naze12}, Paper I). Using this dataset, Naz\'e et al.\ (\cite{naze13}, Paper II) performed an analysis of the variability of the X-ray emission, finding no significant time variability of the line intensities and profiles. Therefore, we can combine the individual RGS spectra into a single, high-quality, fluxed spectrum. The calibration of the fluxed spectrum properly accounts for the dependence of the effective area as a function of energy, but does not account for the redistribution of monochromatic response into the dispersion channels. For intrinsically narrow lines, this might introduce artifacts in spectral regions of strong instrumental sensitivity changes. However, in the case of $\zeta$\,Pup, the intrinsic width of the lines is large and therefore such effects should be of very minor impact here. The instrumental width of the RGS is accounted for in our fitting process, as the synthetic spectra are broadened according to the instrumental profile.

 Whilst the {\it Chandra}-HETG spectrum of $\zeta$\,Pup (Cassinelli et al.\,\cite{cassinelli01}) has a superior spectral resolution and extends to shorter wavelengths than the RGS spectrum analysed in the present work, it has a considerably lower signal-to-noise ratio than the RGS spectrum. We therefore decided to focus our analysis on the RGS spectrum only.

\section{X-ray spectral modeling code}\label{Code}
\subsection{Basic principles of the code}
In order to analyse high-resolution X-ray spectra of massive stars, we have developed our own dedicated code (Herv\'e et al.\ \cite{herve12}). In our model, we assume that the wind hosts several hot plasma components. Each plasma component extends between an inner and an outer boundary and has a specific $kT$ which is constant over that region. The chemical composition is taken to be the same for all components. We then discretize the X-ray emission region of a given plasma component into a large number of cells. We compute the X-ray emissivity of each cell as a function of its temperature and the quantity of material inside the cell. For this purpose we use version 2.0.1 of the AtomDB database (Foster et al.\ \cite{Foster}) with emissivities computed using the APEC code (Smith et al.\ \cite{smith2001})\footnote{The interested reader is invited to visit the section 'physics' for a precise description of the calculation of the continuum and lines in a hot plasma on the AtomDB website at the url http://www.atomdb.org/.}. This code calculates the line emission arising from many physical processes such as collisional excitation, ionization, and dielectronic recombination in a collisionally ionized, optically thin isothermal plasma in thermal equilibrium. The continuum emission is also included under the form of the bremsstrahlung emission and the two-photon decay. 

We assume the plasma to be embedded in the wind and to move along with the latter. The velocity of the wind is determined by a $\beta$-law i.e. $v(r) = v_{\infty} (1-\frac{R_{*}}{r})^{\beta}$ where $v_{\infty}$ is the terminal velocity of the wind and $R_{*}$ is the radius of the star. For each cell, we Doppler-shift its contribution into the observer's frame of reference accounting for the projection of the wind velocity on the line of sight (MacFarlane et al.\ \cite{MacFarlane}).   

Since the X-ray emitting plasma is assumed to be optically thin, an X-ray photon, once emitted, either escapes or is absorbed by the cool wind material along the line of sight. Hence, the radiative transfer problem reduces to the calculation of the optical depth of the fragmented cool wind material, $\tau$, along the line of sight from the emitting cell to the observer. For this purpose, we consider that, at the microscopic level, the cross section for photoelectric absorption in the X-ray domain is dominated by the bound-free transitions from the K and L energy levels.

If the clumps are optically thick, the effective opacity is determined by their geometrical properties (Feldmeier et al.\ \cite{Feld03}). The effective opacity can be written (Owocki \& Cohen \cite{OC06}):
$$\kappa_{eff} = \frac{l^{2}} {m_{c}} =\frac{\kappa}{\tau _{c}}$$
where $l$ is the scale-length of the clump, $m_{c}$ is its mass, $\kappa$ is the mass absorption coefficient and $\tau_{c}$ the optical depth of a clump. This latter can be written:
\begin{equation}
\tau _{c} =\kappa \langle\rho\rangle \frac{l}{f}=\frac{\kappa \dot{M}}{4\pi r^{2}}\frac{1}{v}\frac{l}{f}
\end{equation}
where $\langle\rho\rangle$ is the mean density of the stellar wind at position $r$, $f$ the volume filling factor of the clumps, $\dot{M}$ is the mass-loss rate of the star, and $v$ is the velocity of the wind at the radius $r$ which is determined by a $\beta$-law (see above).

Under this assumption, it appears that the effective reduction of the optical depth as a result of the fragmentation depends on the ratio between the clump scale and filling factor, also called the {\it porosity length} (Owocki \& Cohen \cite{OC06}).
This result can be generalized from the optically thick to the optically thin limit using a simplified scaled effective opacity (Owocki \& Cohen \cite{OC06}): 
\begin{equation}
\frac{\kappa_{eff}}{\kappa}=\frac{1}{1+\tau_{c}} 
\label{eqn2}
\end{equation}
Using this result in a steady state (i.e.\ non variable) wind, we obtain the wind optical depth (Owocki \& Cohen \cite{OC06}) in conventional $(p,z)$ coordinates:

\begin{equation}
\tau_{\lambda}(p,z)=\tau_{\ast} \int^{\infty}_{z} \frac{R_{\ast}dz'}{r'(r'-R_{\ast})+\tau_{\ast}\,R_{\ast}\,h(r')}
\label{eqowo}
\end{equation}
where $r=\sqrt{p^{2}+z^{2}}$, $\tau_{\ast}\equiv \frac{\kappa \dot{M}}{4\pi v_{\infty}R_{\ast}}$ is the characteristic optical depth of the wind, and $h(r)\equiv\frac{l}{f}$ is the porosity length. In the following we adopt $h(r)=h'\times r$.

A slightly different formalism to describe the fragmentation of the wind is based on the concept of the {\it{fragmentation frequency}} (Oskinova et al.\ \cite{OFH}). As we have shown in Herv\'e et al.\ (\cite{herve12}), both formalisms yield very similar X-ray line profiles and the capabilities of the current generation of high-resolution X-ray spectrographs aboard {\it XMM-Newton} or {\it Chandra} do not allow us to choose between the two formalisms. Here we adopt the porosity length formalism because it leads to an analytic expression for the integral in the equation of the optical depth. This has the advantage to allow a faster computation. In Herv\'e et al.\ (\cite{herve12}), we have shown that the inclusion of a few percent of the wind material in a homogeneous inter-clump medium has little impact on the emerging X-ray spectrum. In the present work, we thus consider that the inter-clump medium is void. 

In either formalism, the characteristic wind optical depth $\tau_*$, and thus the mass absorption coefficient $\kappa$, play a key role in the computation of the absorption of X-rays by the stellar wind. To estimate the values of $\kappa$, we use the non-LTE, line blanketed, model atmosphere code CMFGEN (Hillier \& Miller \cite{HillierMiller98}) to compute the ionization structure of the cool wind as a function of temperature and mass-loss rate. In these ionization calculations, CMFGEN accounts for  different physical processes such as collisional excitation, photo-ionization by photospheric light and X-rays, as well as radiative and dielectronic recombination. Once we know the ionization structure, we can then compute $\kappa$. 

\begin{figure}
\begin{center}
\includegraphics[width=4.25cm]{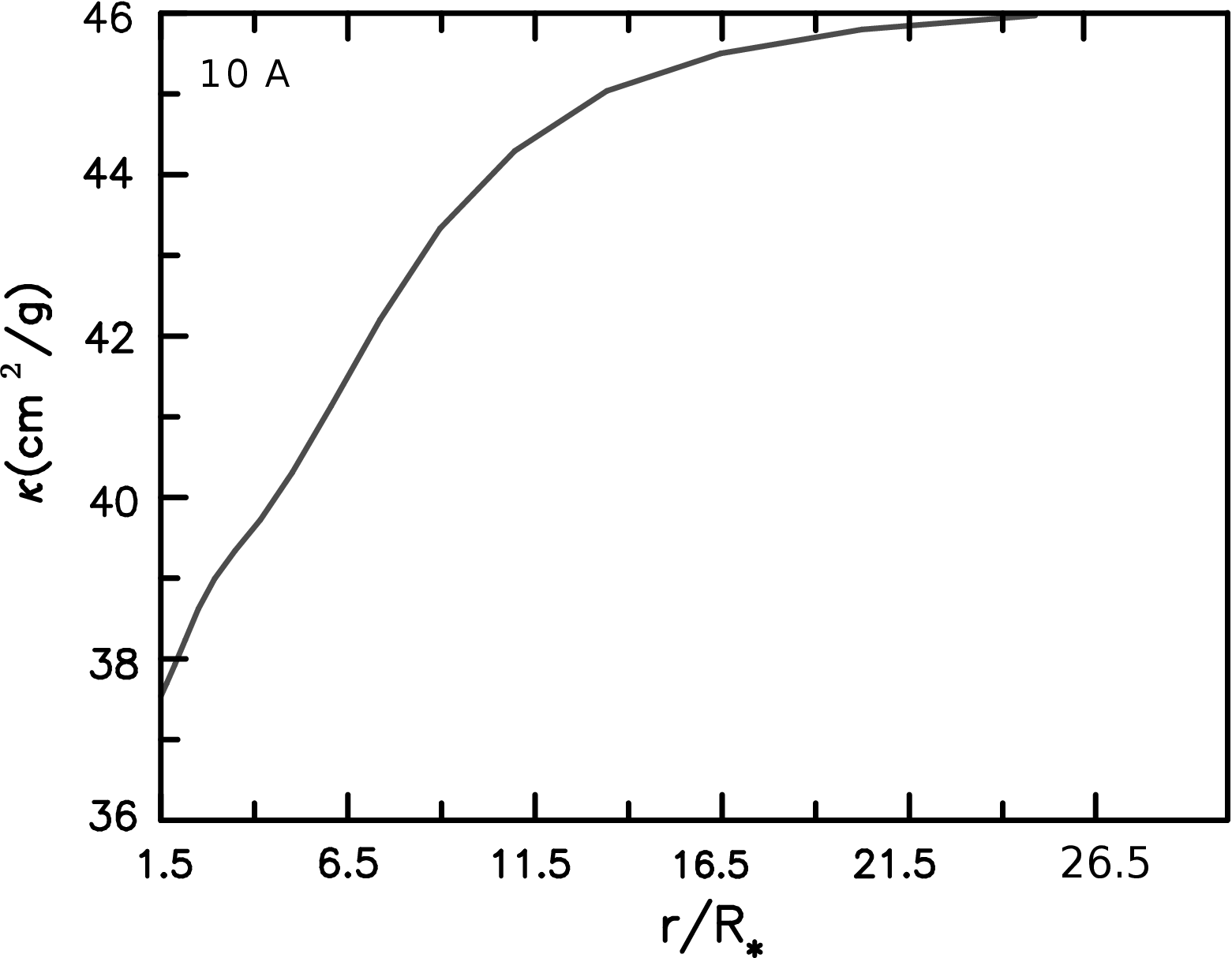}
\includegraphics[width=4.25cm]{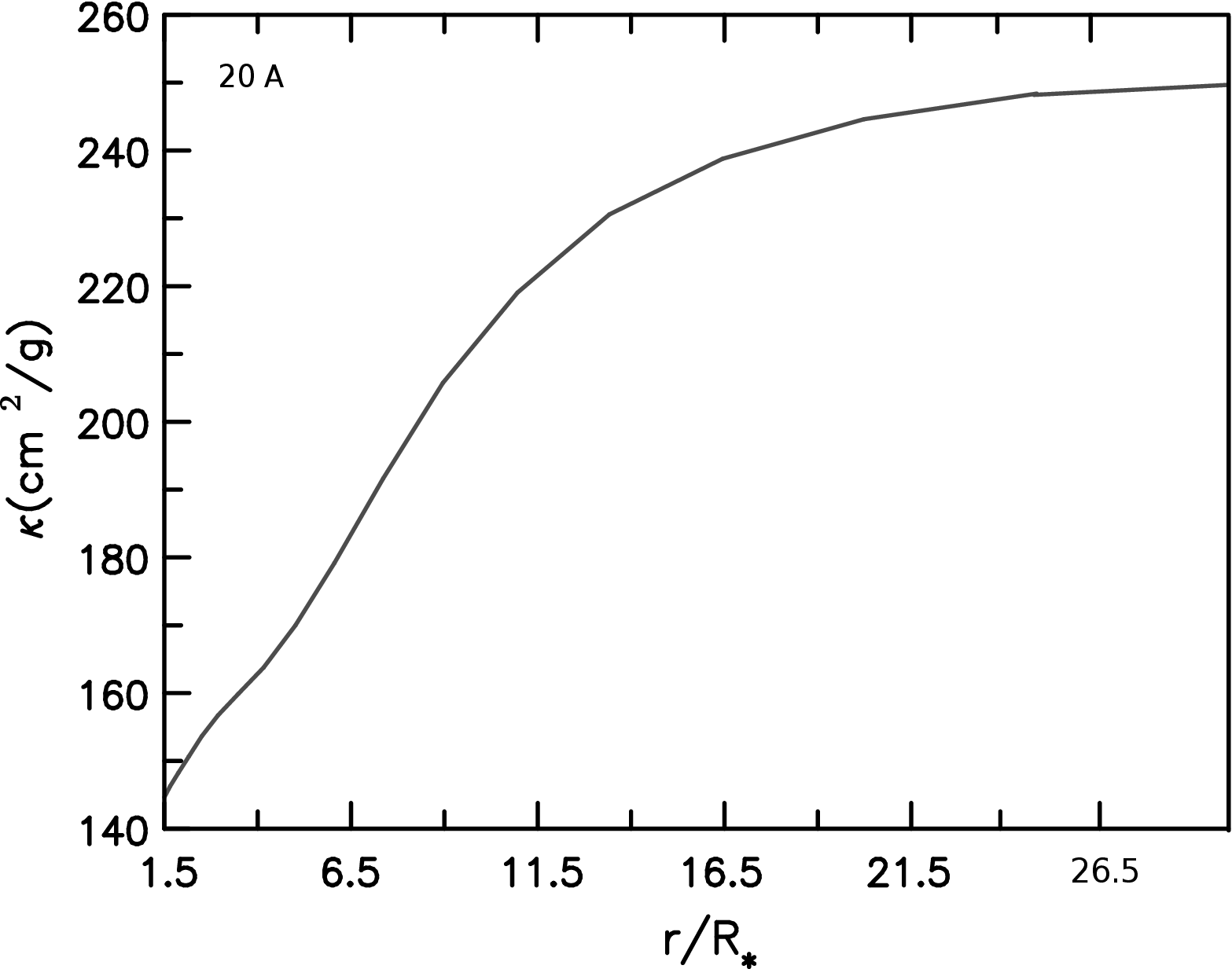}
\includegraphics[width=4.25cm]{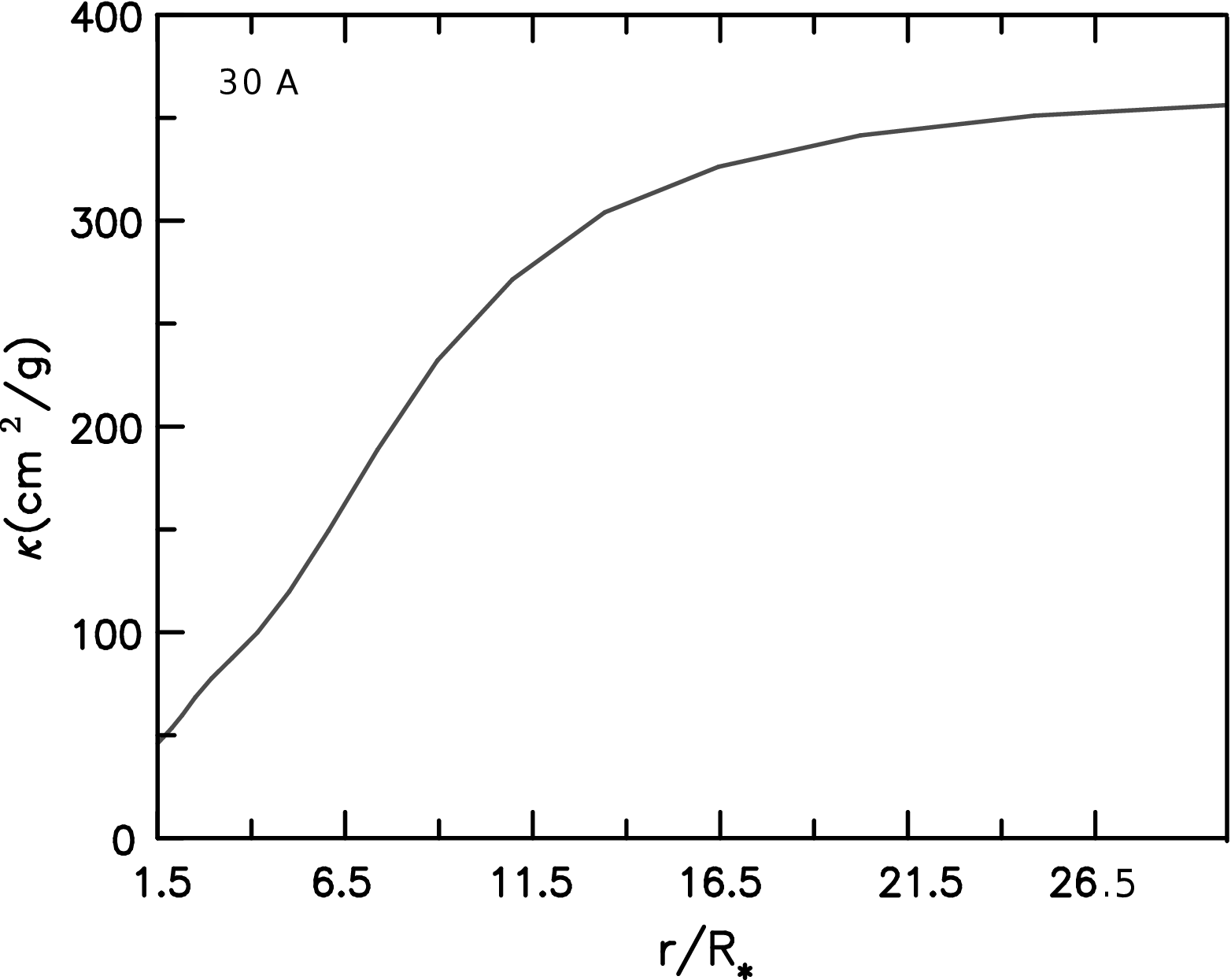}
\includegraphics[width=4.25cm]{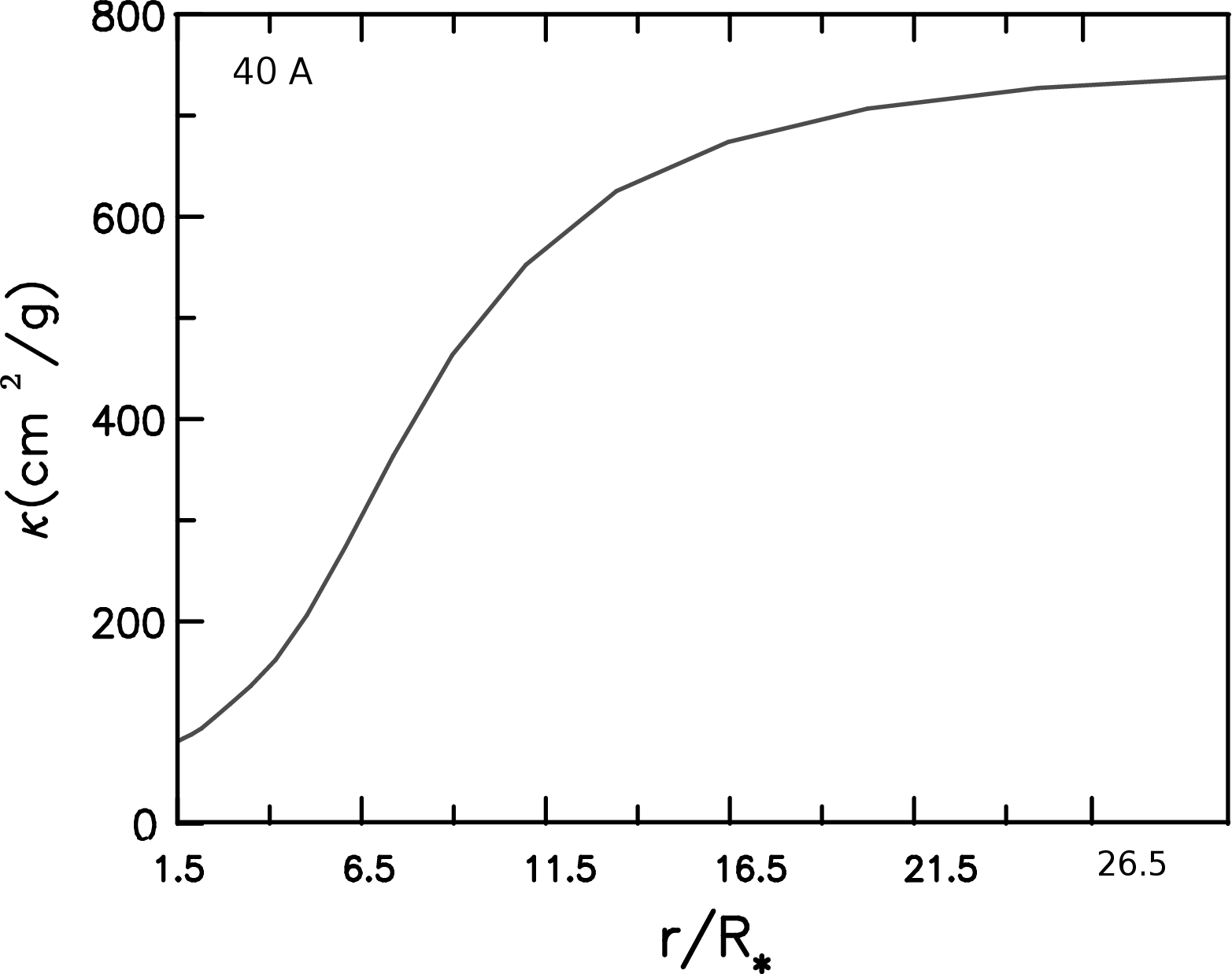}
\caption{Radial dependence of the mass absorption coefficient at different wavelengths (specified in the upper left corner of each panel). The mass absorption coefficients are calculated with the elements listed in Tab.\ref{tab1} with abundances found in our study.}
\label{fig1}
\end{center}

\end{figure}

The results are illustrated in Fig.\,\ref{fig1} at four different wavelengths. This plot shows that the approximation of a mass absorption coefficient that is independent of the position in the wind is not a valid one in this case. Indeed, the ionization structure of the wind is not constant throughout the entire wind and the impact on the value of the mass absorption coefficient is huge. The outer regions of the wind are less ionized and consequently, the cool material produces a stronger absorption in the outer regions. 

\begin{table}
\caption{Chemical elements and their ionization levels used to calculate the mass absorption coefficient $\kappa$ and the UV radiation field in CMFGEN.}
\begin{center}
\begin{tabular}{ll}
\hline
\hline
H & I \\
He & I,II \\
C  & II,III,IV\\
N & II,III,IV,V\\
O & II,III,IV\\
Mg & II,III,IV \\
Si & III,IV\\
Ar & III, IV,V \\
Fe & III,IV,V,VI,VII\\
Ni & III,IV,V \\
P & IV,V \\
Al & III \\
Na & I \\
\hline
\hline
\end{tabular}
\end{center}
\label{tab1}
\end{table}

\begin{figure}
\begin{center}
\includegraphics[width=4.25cm]{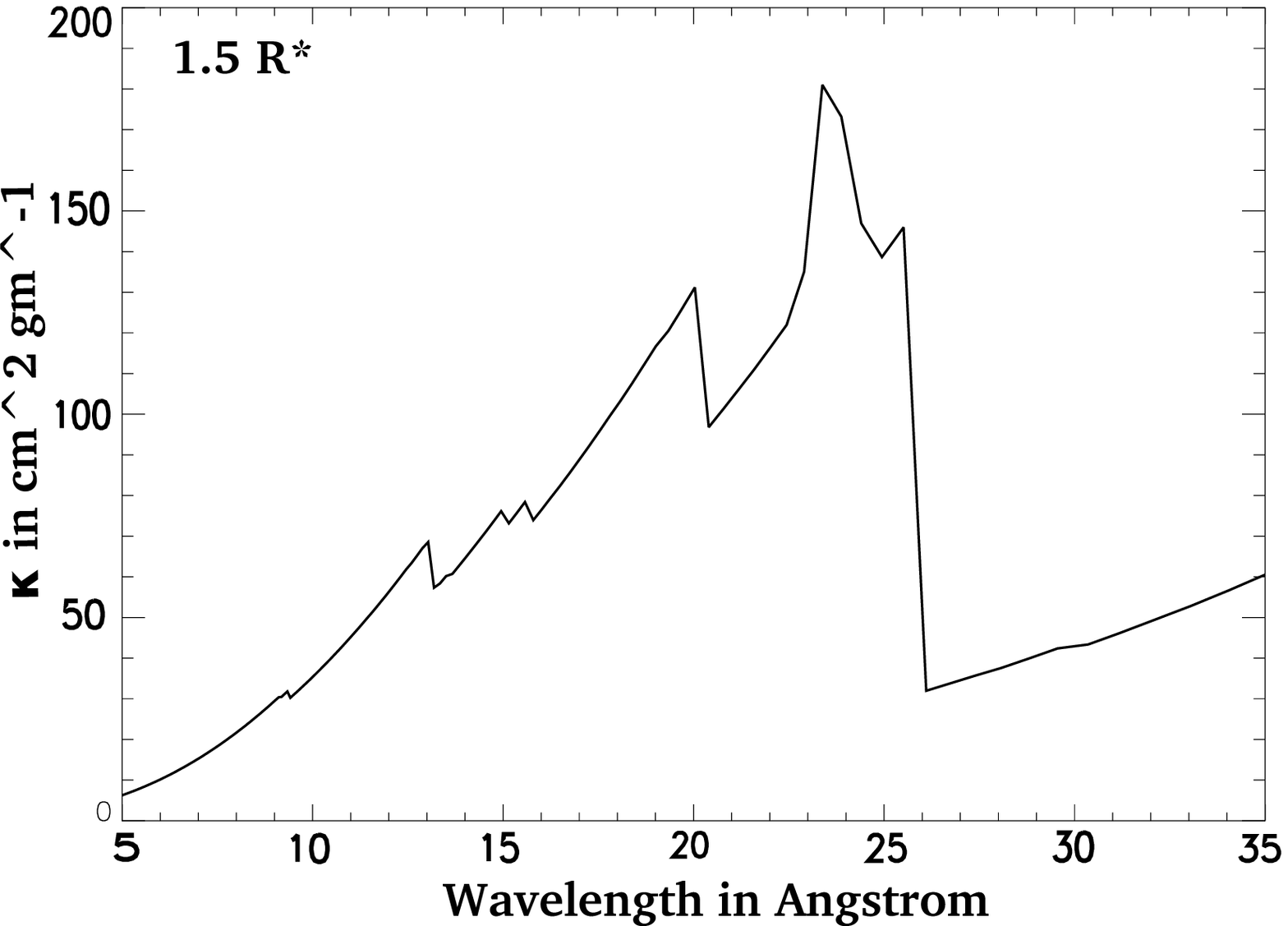}
\includegraphics[width=4.25cm]{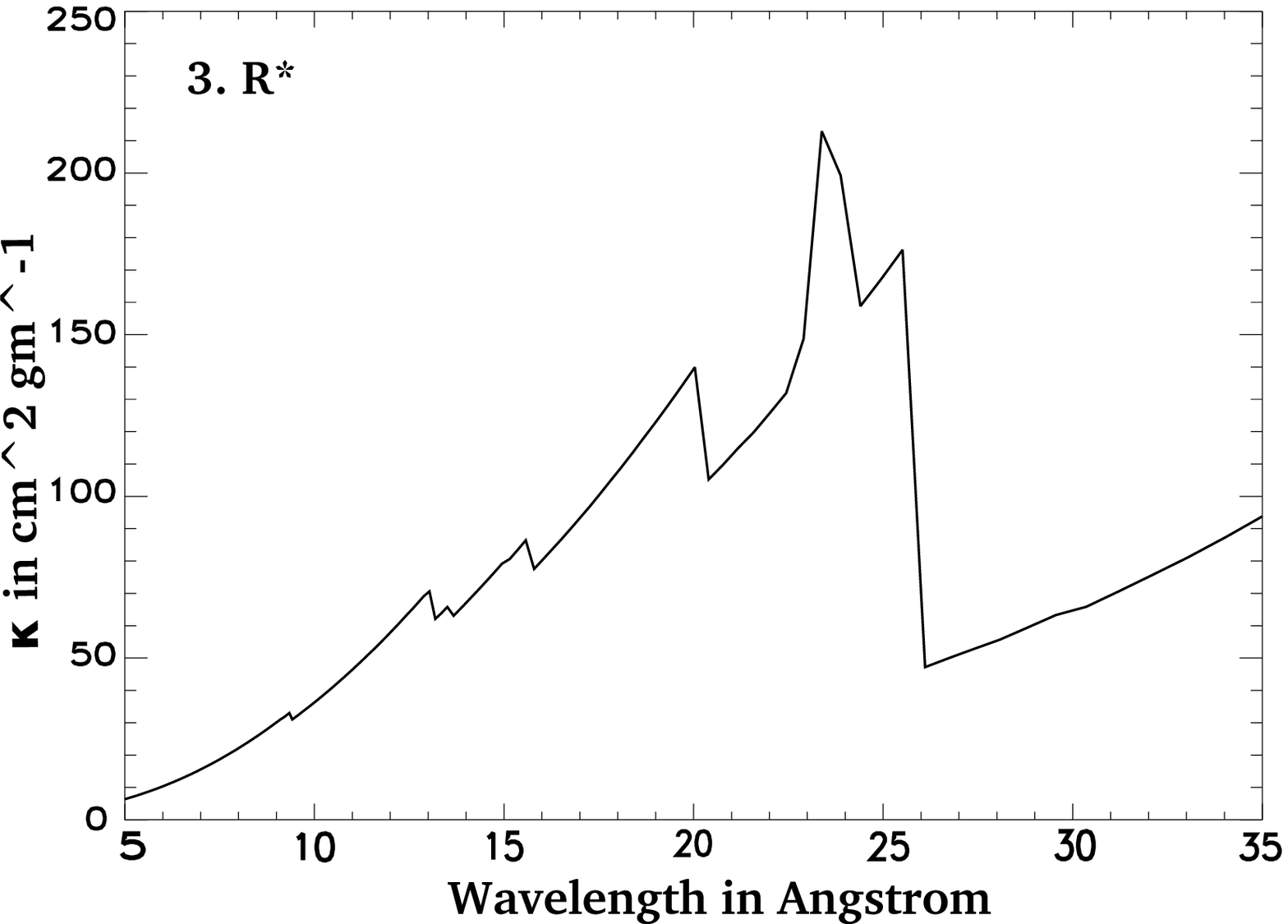}
\includegraphics[width=4.25cm]{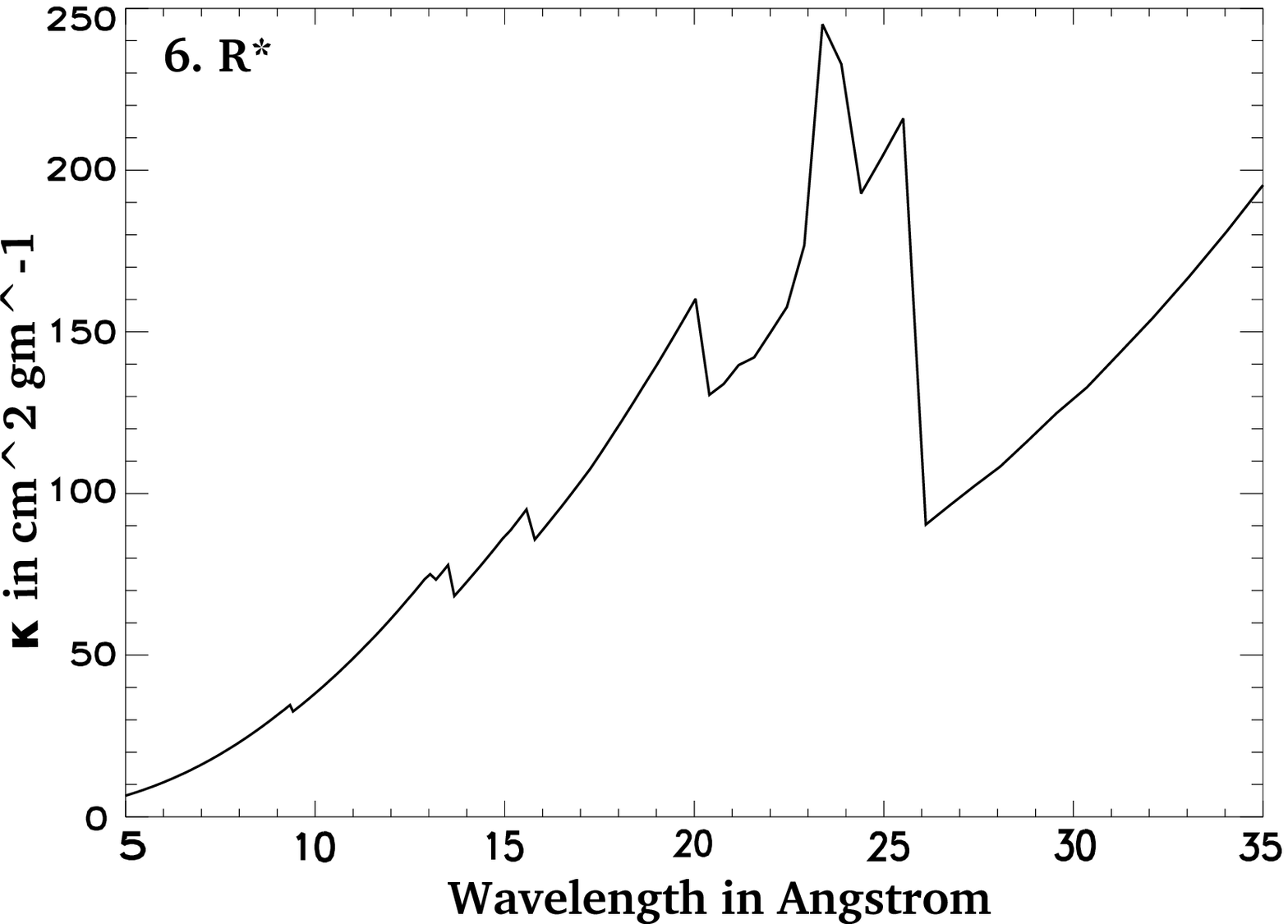}
\includegraphics[width=4.25cm]{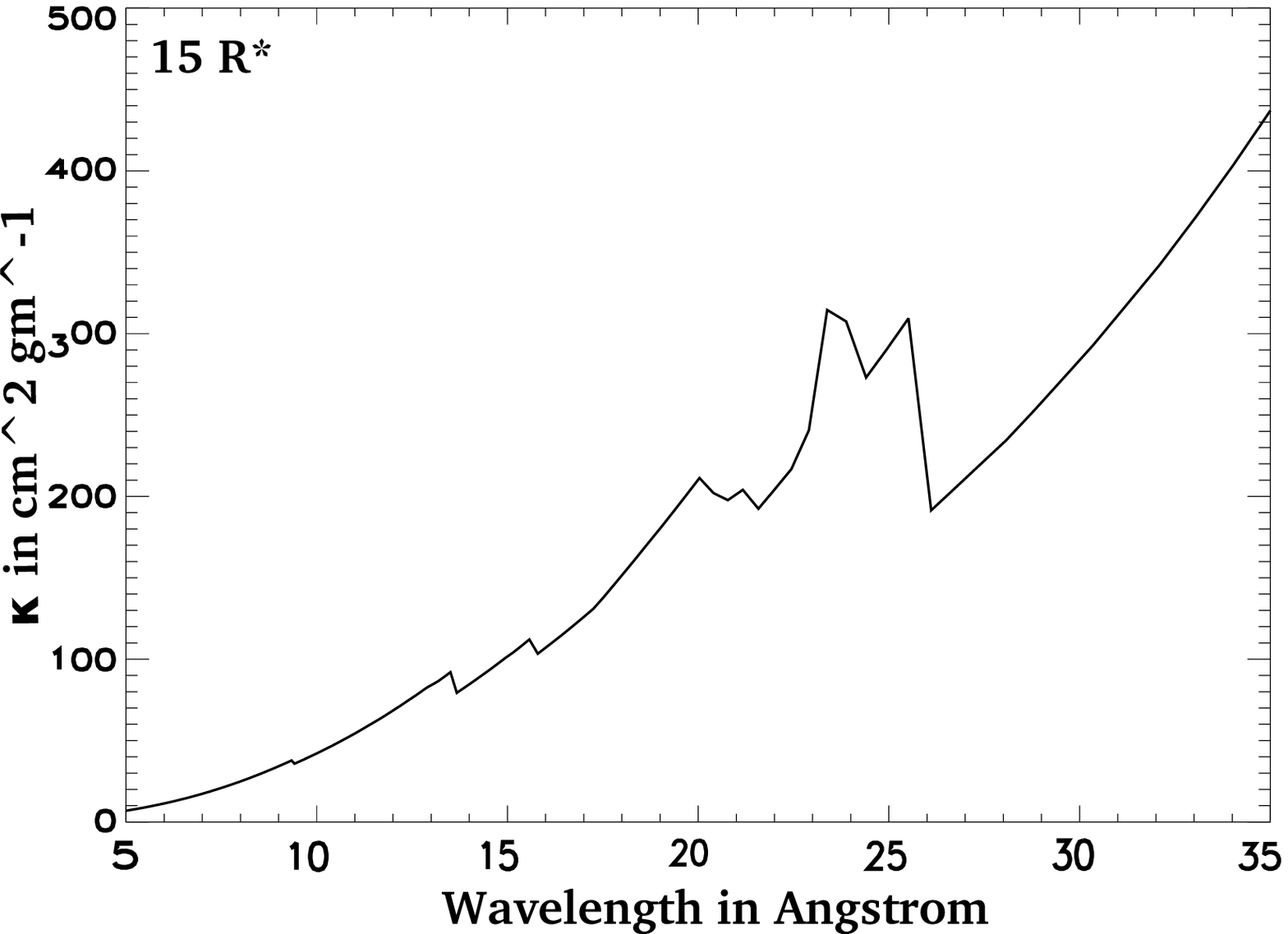}
\caption{Wavelength dependence of the mass absorption coefficient at different positions inside the wind ($1.5$, $3$, $6$ and $15\,R_*$ as specified in the upper left corner of each panel).}
\label{fig1bis}
\end{center}
\end{figure}

Compared to our previous version of the code (Herv\'e et al.\ \cite{herve12}) where we adopted a value of $\kappa$ that was independent of $r$, we have decided to render the treatment of the absorption of the wind more realistic. For this purpose, we approximate the radial dependence of $\kappa$ by two simple relations that hold over two ranges of values of $r$. A linear relation 
\begin{equation}
\kappa (r) = \kappa_{0} + \gamma (r - R_{*})
\label{Eq4}
\end{equation}
for $r < R_{lim}$, and a constant 
\begin{equation}
\kappa (r) = \kappa_{max}
\end{equation}
for $r \geq R_{lim}$. Here $\kappa_{0}$ and $\kappa_{max}$ are, respectively, the mass absorption coefficient in the innermost part and in the outer part of the wind and $\gamma$ is the slope. $R_{lim}$ is the position in the wind where $\kappa$ reaches its plateau value $\kappa_{max}$ (see Fig.\,\ref{fig1}). The advantage of this parametrization is to provide a more realistic description of the variation of $\kappa$ as a function of the radius and, at the same time, to preserve an analytical solution for the integral of the optical depth equation.
Indeed, for a linearly increasing porosity length ($h(r) = h' \times r$), the optical depth is now given by:
\begin{eqnarray} 
\tau_{\lambda}(p,z) & = & \int_{z}^{\sqrt{(R_{lim})^2 - p^2}} \!\frac{\alpha\,(\kappa_{0} - \gamma\,R_{*})\,R_{*}\,dz'}{r'\,(r'-R_{*}) + \alpha\,(\kappa_{0} + \gamma\,(r'- R_{*}))\,R_{*}\,h'\,r'} \nonumber \\
& +  & \int_{z}^{\sqrt{(R_{lim})^2 - p^2}} \! \frac{\alpha\,\gamma\,R_{*}\,dz'}{(r'-R_{*}) + \alpha\,(\kappa_{0} + \gamma\,(r'- R_{*}))\,R_{*}\,h'} \nonumber \\
&  + & \int_{\sqrt{(R_{lim})^2 - p^2}}^{\infty} \! \frac{\alpha\,\kappa_{max}\,R_{*}\,dz'}{r'\,(r'-R_{*}) + \alpha\,\kappa_{max}\,R_{*}\,h'\,r'} 
\label{taufinal}
\end{eqnarray}
with $\alpha = \frac{\dot{M}}{4\pi R_{*}v_{\infty}}$. 

Note that we have assumed in Eq.\,\ref{taufinal} that $z$ is included in the interval $[-\sqrt{R^2_{lim} - p^2}, \sqrt{R^2_{lim} - p^2}]$. If the latter condition is not fulfilled, the boundaries of these integrals are adjusted, but the principle remains the same.

The first two integrals have analytical solutions making the computation easier and the same is true for the last integral which is formally identical to the one used in the original prescription of Owocki \& Cohen (\cite{OC06}, see our Eq.\,\ref{eqowo}).

Figure\,\ref{fig1bis} illustrates the wavelength dependence of the mass absorption coefficient at four different positions in the stellar wind. The strongest absorption edge in this figure (at about 26\,\AA) is due to N\,{\sc iv}. The complex shape of this relation could impact some of the frequently used diagnostic tools based on line ratios. For instance, the ratio between the flux of N\,{\sc vii} Ly\,$\alpha$ at 24.78\,\AA\ and the flux of the N\,{\sc vi} triplets at 28.8 -- 29.5\,\AA, which is used in line-by-line analyses as a temperature indicator, could be affected by the strong N\,{\sc iv} edge.

\subsection{The treatment of the {\it fir} triplets of He-like ions}
In an optically thin thermal plasma at high temperatures, some elements retain only two electrons and are thus present as ions with a He-like electron configuration (e.g.\ O\,{\sc vii} and N\,{\sc vi}). Such ions produce a triplet of lines consisting of so-called forbidden 2\,$^3S_1$ - 2\,$^1S_0$ ({\it f}), inter-combination 2\,$^3P_{1,2}$ - 2\,$^1S_0$ ({\it i}) and resonance 2\,$^1P_{1}$ - 2\,$^1S_0$({\it r}) lines (Gabriel \& Jordan \cite{GJ}). In low density environments or in the absence of strong UV radiation fields, the upper level of the {\it f} transition is only depopulated via the forbidden line which is then seen as a rather strong line in the spectra (Blumenthal et al.\ \cite{blue71}). However, in the wind of a massive star, the presence of an intense UV radiation field alters this situation. Indeed, the UV photons pump the electrons from the upper level of the {\it f} transition to the upper level of the {\it i} line (Blumenthal et al.\ \cite{blue71}, Porquet et al.\ \cite{porquet01}). Therefore the ratio, ${\cal R}=\frac{f}{i}$ is strongly modified which  provides an interesting tool to constrain the dilution of the UV radiation field at the position of the X-ray emission and hence to determine the location of the hot plasma in the wind.

Since the flux of UV radiation seen by the ions in the stellar wind decreases in proportion to the geometrical dilution factor $w(r)=\frac{1}{2} (1 - [1- (\frac{R_{*}}{r})^2]^{\frac{1}{2}})$, the ${\cal R}$ ratio is also a function of radius. The radial modification due to the photoexcitation from the metastable upper level of the forbidden line can be expressed as follows (Blumenthal et al.\ \cite{blue71}, Leutenegger et al.\ \cite{Leutenegger06}): 
\begin{equation}
{\cal R}(r) = {\cal R}_{0} \frac{1}{1 + 2\,P\,w(r) + N_e/N_c}
\end{equation}
where $P=\frac{\phi_{*}}{\phi_{c}}$ and $\phi_{*} = 8\pi \frac{\pi e^{2}}{m_{e}v}f\frac{H_{\nu}}{h\,\nu}$
with $H_{\nu}$ the Eddington flux and $f$ is the sum of the oscillator strengths for 2\,${^3}S_{1}$ to all three of 2\,$^{3}P_{J}$. $N_e$ is the free electron density. $\phi_{c}$ and $N_c$ are the critical photoexcitation rate and the critical density (which depend on atomic parameters and the electron temperature, see Blumenthal et al.\ \cite{blue71}). In the winds of hot, luminous massive stars, the photoexcitation rate dominates over the collisional excitation term, and in the remainder of this work, we have thus neglected the $N_e/N_c$ term. 

The treatment of these effects is now included in our code along the above prescriptions and using the Eddington fluxes computed with our CMFGEN model. In Table\,\ref{tab3}, we indicate the physical information necessary to reproduce this modification of intensities of the forbidden and inter-combination line present in the spectra of $\zeta$\,Pup. 

\begin{table*}
\caption{Parameters for the treatment of the forbidden, inter-combination and resonance lines.}
\begin{center}
\begin{tabular}{cccccccccc}
\hline 
Ion  & $\lambda_{r}$ & $\lambda_{i}$ & $\lambda_{f}$ & $\lambda_{uv(i1)}$ & $\lambda_{uv(i2)}$ & $\phi_{c}$ & $H_{\nu 1}/(h\,\nu)$& $H_{\nu 2}/(h\,\nu)$ & $f$\\
\hline
N\,{\sc vi} & 28.792  & 29.074,  29.076 & 29.531 & 1907.26 & 1896.74 & $1.83 \times 10^2$ & 1.71  & 2.27 & 0.1136\\
\hline
O\,{\sc vii} &  21.603 & 21.796, 21.799 & 22.095 & 1638.28 & 1623.61 & $7.32 \times 10^2$ & 1.65  & 1.30 & 0.0975\\
\hline
Ne\,{\sc ix} &  13.447 & 13.548, 13.551 & 13.697 & 1272.81 & 1248.28 & $7.73 \times 10^3$ & 1.95 & 1.87 &0.0700\\
\hline
Mg\,{\sc xi} &  9.1681 & 9.2267, 9.2298 & 9.3134 & 1034.31 & 997.46 & $4.86 \times 10^4$ & 1.19 & 1.10 &0.0647\\
\hline
Si\,{\sc xiii} &  6.6471 & 6.6838, 6.6869 & 6.7394 & 865.14 & 814.69 & $2.39 \times 10^5$ & 0.64 & 0.86 &0.0562\\
\hline

\end{tabular}
\end{center}
\label{tab3}
\tablefoot{$\lambda_{uv(i1)}$ and $\lambda_{uv(i2)}$ correspond to the UV radiation wavelengths which depopulate the inter-combination level of a given element. All wavelengths are given in \AA. \\
$\phi_{c}$ are taken from Blumenthal et al.\ (\cite{blue71}). \\
$H_{\nu}/(h\,\nu)$ are calculated with CMFGEN with our own abundances and mass-loss rate but with $T_{eff}$ determined in the UV/optical domain. The values are given in $10^{8}$ photon cm$^{-2}$s$^{-1}$Hz$^{-1}$. \\
The oscillator strengths, $f$, are extracted from the AtomDB website.}
\end{table*}

\section{Multi-temperature plasma fitting procedure}\label{multiT}
The full X-ray spectrum of a massive star can usually not be fitted with a single temperature plasma model. In the present case for instance, the most prominent ionization level of iron (Fe\,{\sc xvii}) seen in the RGS spectrum of $\zeta$\,Pup, suggests the presence of a hot plasma component with a $kT$ close to  0.50\,keV. Other lines, such as the Si\,{\sc xiii} lines around 6.7\,\AA, have their maximum emissivity at even higher temperatures around 0.90\,keV. On the other hand, the {\it fir} triplets of He-like carbon and nitrogen are rather indicative of a lower temperature plasma ($kT$ around 0.10 keV). Fitting the observed spectrum hence requires the combination of several models with different temperatures. At this stage, we have to stress that the HETG spectrum of $\zeta$\,Pup (Cassinelli et al.\ \cite{cassinelli01}) as well as its EPIC spectra (Naz\'e et al.\ \cite{naze12}) reveal the presence of a weak Si\,{\sc xiv} Ly\,$\alpha$ line at 6.18\,\AA\ and of a weak S\,{\sc xv} triplet around 5.04 -- 5.10\,\AA. These ions have their strongest emission at temperatures near 1.3 -- 1.4\,keV. However, these lines are outside the sensitivity range of the RGS instrument and are thus not included in our fit. We will return to this point in Sect.\,\ref{Results}.

Of course, each line is not only emitted at the temperature where its emissivity reaches its maximum value, but rather exists over some range of temperatures, although with a lower intensity. Therefore most lines in the observed spectrum result from the combination of several contributions associated with plasma components at different temperatures (see Fig.\,\ref{fig3}). The main difficulty in the fitting procedure therefore consists in finding the best values of the parameters that describe our synthetic spectra, including the filling factor $f_{hot gas}$  (i.e the ratio of the X-ray emitting volume and the total volume) for each plasma component.  

We have thus developed a code which combines different components at different temperatures in order to reproduce the total spectrum of $\zeta$\,Pup by a minimization of the $\chi^{2}$ of the fit.
 We have based our fitting procedure on a multi regression routine\footnote{We use the IDL routine IMSL\_Multiregress. More information are available at http://idlastro.gsfc.nasa.gov/idl$\_$html$\_$help/IMSL$\_$Multiregress.html}. 
 There are quite a number of parameters needed to describe a model. These are the general wind and stellar parameters ($\dot{M}$, $v_{\infty}$, $\beta$, $R_*$ and $T_*$), the porosity parameter $h'$, the abundances of the different elements that make up the hot plasma, and for each plasma component, its temperature ($kT$), filling factor ($f_{hot gas}$), and inner and outer boundary (R$_{in}$ and R$_{out}$). Not all of them are treated as free parameters in the present study: only the CNO abundances are varied with respect to solar; $v_{\infty}$, $R_*$ and $T_*$ are all taken from Bouret et al.\ (\cite{bouret12}), whilst we adopt $\beta = 1$ to preserve the analytical solution of the optical depth integrals. Nevertheless, we are left with $4 \times n + 5$ parameters where $n$ is the number of plasma components. In our case, we find that $n = 4$ is required to fit the observed spectrum and we are thus dealing with 21 free parameters. In order to obtain the best solution we used our synthetic spectra simulator code (Herv\'e et al.\ \cite{herve12}) to build a database of individual plasma components (hence nine free parameters per element of the database). In doing so, each parameter is sampled  over a given range that we consider reasonable (e.g.\ $kT$ varies between 0.08 and 0.80\,keV and $\dot{M}$ varies in the range $1.75 \times 10^{-6}$ to $4.0 \times 10^{-6}$\,$M_{\odot}$\,yr$^{-1}$).

 The fitting routine then systematically explores this database, combining only plasma components that have the same chemical composition, $h'$ and $\dot{M}$. The synthetic multi-temperature spectrum is built as follows:
\begin{equation}
S_{i} = \sum_{j=1}^n f_{hotgas, j} M_{j,i}  + \epsilon
\end{equation}
where $S_{i}$ is the observed flux at the wavelength $i$, $f_{hotgas, j}$ is the classical volume filling factor of the hot gas component $j$, $M_{j,i}$ is the synthetic flux of each X-ray emitting plasma component at the wavelength $i$, and $\epsilon$ is the error of regression. 

For each combination of plasma models, the best fit values of $f_{hotgas, j}$ are determined in the least-squares sense and the resulting $\chi^2_{\nu}$ is obtained comparing the model to the fluxed spectrum. These numbers are stored for later use, including the identification of the best-fit model based on the value of $\chi^2_{\nu}$. After several tests, it appears that at least four components are needed to achieve a good fit of the fluxed RGS spectrum of $\zeta$\,Pup.

We have included an additional physical constraint to this fitting procedure : we imposed the global X-ray emitting region around the star to be continuous. In other words, from the lowest inner boundary radius of the emitting regions to the largest outer boundary radius, there must not exist any spatial gap where there is no hot plasma emission at all. 

It is difficult to make any statement about the unicity of the solution. There might indeed exist other combinations of (more than four) plasma components with values of the wind parameters ($\dot{M}$, $h'$,...) outside the range explored here that could fit the spectrum. However, within the assumptions made here (four discrete plasma components,...), we are confident that our procedure has identified the best solutions.

\section{Results and discussion}\label{Results}
\begin{figure*}[htb]
\begin{center}
\includegraphics[width=14cm]{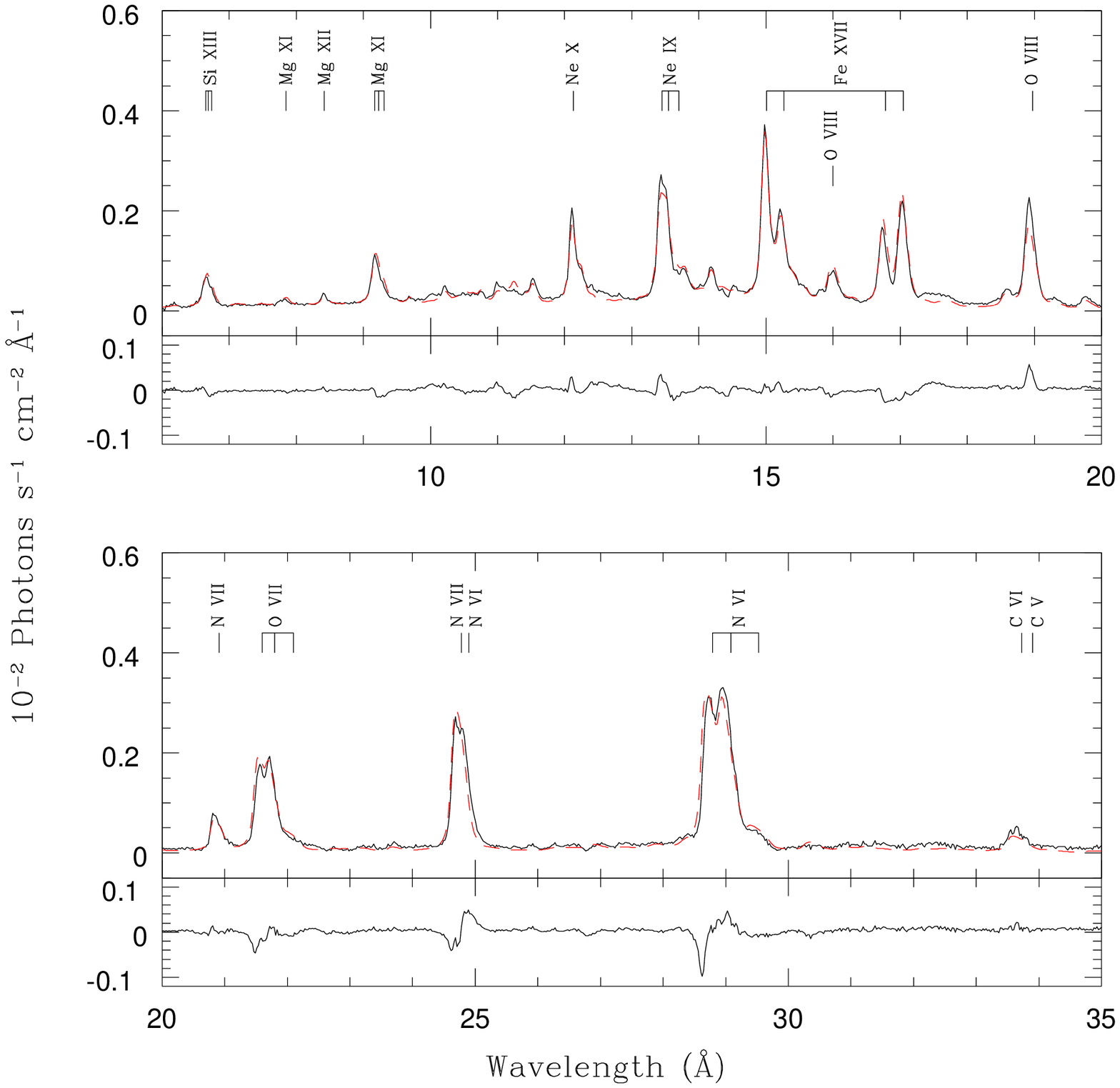} 
\caption{ Our best-fit model (model$_1$ of Table\,\ref{tab4}, in dashed red line) of the observed RGS spectrum of $\zeta$\,Pup (in solid black line). The most prominent emission lines are labelled. We note that there are many more weaker lines that contribute to the spectrum and are included in our model. The residuals (in the sense observation minus model) are shown in the panels below the spectrum.}
\label{fig2}
\end{center}
\end{figure*}

To start our analysis, we used the stellar parameters of Bouret et al.\ (\cite{bouret12}). Then we explore a large range of values for all the free parameters in order to minimize the $\chi^{2}$ of the fit. Our best-fit model of the full RGS spectrum of $\zeta$\,Pup is illustrated in Fig.\,\ref{fig2}. 

\subsection{Temperature distribution and hot gas filling factor}

\begin{table*}
\caption{Best-fit X-ray emitting plasma parameters found in our study.}
\begin{center}
\begin{tabular}{c|ccc|ccc}
\hline
 & model$_1$ & h'=0.0 & $\chi^{2}_{\nu}$ = 15.16 & model$_2$& h'=0.02 & $\chi^{2}_{\nu}$ = 17.82 \\
\hline
$kT$  & $f_{hot gas}$ & R$_{in}$  & R$_{out}$ & $f_{hot gas}$ & R$_{in}$  & R$_{out}$\\
 keV & &  $R_{*}$& $R_{*}$ &  &  $R_{*}$& $R_{*}$\\
\hline
0.10&  0.012 & 7.5 & 85. &  0.013 & 2.8 & 95. \\
\hline
0.20&  0.012 & 1.5 & 38. &  0.012 & 1.6 & 40. \\
\hline
0.40&  0.020 & 2.7 & 4.0 &  0.020 & 3.0 & 4.2  \\
\hline
0.69&  0.007 & 3.1 & 4.1 &  0.007 & 3.2 & 4.9 \\
\hline
\end{tabular}
\end{center}
\label{tab4}
\tablefoot{ $\chi^{2}$ is the reduced chi-squared with 1222 degrees of freedom.}
\end{table*}

\begin{figure*}
\begin{center}
\includegraphics[width=10cm,angle=90]{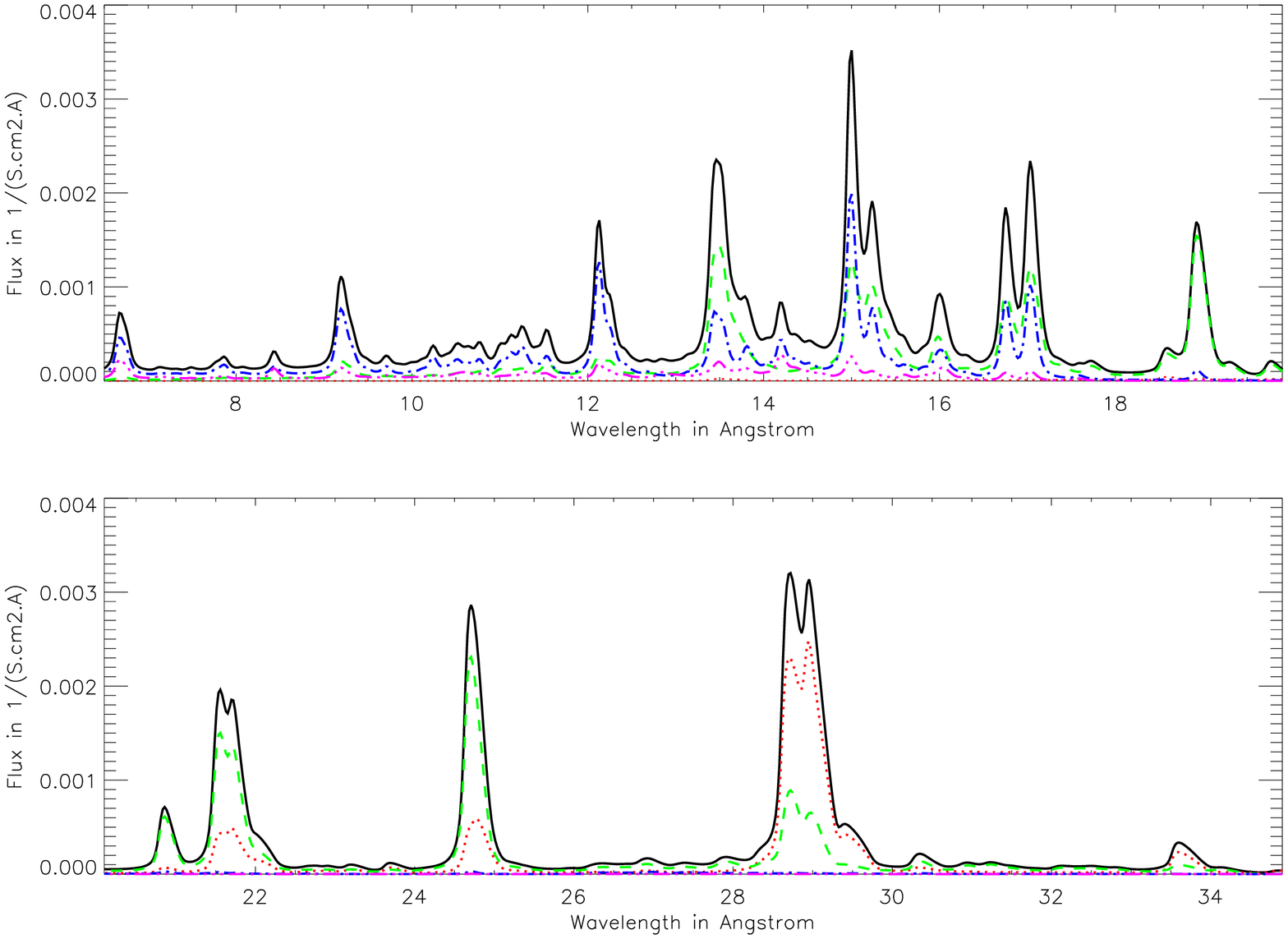} 
\caption{ Contributions of the four plasma components to the best-fit model of the RGS spectrum of $\zeta$\,Pup (model$_1$, in solid black line): $kT_1$ = 0.10\,keV (in dotted red line), $kT_2$ = 0.20\,keV (in dashed green line), $kT_3$ = 0.40\,keV (in dotted dashed blue line) and $kT_4$ = 0.69\,keV( in long dashed dotted pink line).}
\label{fig3}
\end{center}
\end{figure*}

The true temperature distribution of the X-ray emitting plasma in the wind of $\zeta$\,Pup is almost certainly a continuum of temperatures between  a minimum and a maximum temperature. Nevertheless, a discretization with four different temperatures is sufficient to reproduce correctly the observed  RGS spectrum of $\zeta$\,Pup. As pointed out above, these four components might not be sufficient to account for the presence of the highly ionized Si\,{\sc xiv} and S\,{\sc xv} lines seen in the HETG and EPIC spectra. However, although the RGS sensitivity range does not cover the S\,{\sc xv} triplet, our model predicts an integrated line flux of $0.64 \times 10^{-13}$\,erg\,cm$^{-2}$\,s$^{-1}$, which is not too far away from the value $0.81 \times 10^{-13}$\,erg\,cm$^{-2}$\,s$^{-1}$ reported by Cassinelli et al.\ (\cite{cassinelli01}), especially in view of the relative uncertainties on this flux. The latter uncertainties are likely of order 10\%, as this triplet contains between 80 and 100 counts in the HETG spectrum, as judged from the {\it Chandra} transmission grating catalog (Huenemoerder et al.\ \cite{huenemoerder})\footnote{see http://tgcat.mit.edu}. Therefore, if a hotter plasma component is needed to fully account for the flux of this triplet, its emission measure is likely very low.

 In this respect, it is interesting also to compare our results with the fits of the EPIC spectra reported in Paper I (Naz\'e et al.\ \cite{naze12}). In that paper, we used a four-temperature thermal plasma model with an overlying `slab' of absorbing material. The best fit plasma temperatures were found to be $kT_{1,2,3,4} =$ 0.09, 0.27, 0.56 and 2.18\,keV. Whilst the first three temperatures are very similar to - or mean values of - what we find here, the fourth one is clearly much higher than found in our analysis. However, the emission measure of the 2.18\,keV component in the fits of Paper I, is about a factor 100 below that of the 0.56\,keV component. Therefore, this very hot plasma should indeed have a very limited impact on the RGS spectrum analysed here. Nevertheless, this remark implies that there exists very hot material in the wind of $\zeta$\,Pup. It would thus be extremely interesting to collect a HETG spectrum of quality comparable to the RGS spectrum, and to repeat the kind of analysis presented here, to constrain the location of this very hot plasma.

 In previous analyses, the line intensity ratios of hydrogen-like and helium-like ions was used as a diagnostic of the X-ray source temperature (e.g.\ Waldron \& Cassinelli \cite{WC07}). Such ratios are expected to provide some average temperature between the various plasma components that contribute to the line formation. However, Fig.\,\ref{fig3} indicates that the relative weights of the different plasma components in this average depend upon the element that is considered. For instance, in the case of oxygen, the O\,{\sc viii} Ly$\alpha$ line is dominated by the 0.20\,keV plasma, whilst the O\,{\sc vii} triplet includes contributions from the 0.10 and 0.20\,keV plasma components. Waldron \& Cassinelli (\cite{WC07}) derived $kT$ = 0.24\,keV from the oxygen line ratio, which is consistent with the dominant role of the 0.20\,keV plasma in this case. If we consider neon instead, we see that the Ne\,{\sc x} and Ne\,{\sc ix} lines are mostly formed in the 0.40 and 0.20\,keV components, with minor contributions from the 0.69\,keV component. For these ions, Waldron \& Cassinelli (\cite{WC07}) obtained $kT$ = 0.33\,keV from the line ratio. This is consistent with the fact that, in this specific case, the two dominant components have their relative importance reverted in the lines of the H-like and He-like ions. In summary, whilst we confirm that line ratios yield some average estimate of the temperature, it is impossible to know, from a line-by-line analysis, what are the weights of the different plasma components in this average.

 The two plasma components with the lowest temperatures ($kT$ of 0.10 keV and 0.20 keV) extend over the widest range in radius and dominate the emission (model$_1$ in Table\,\ref{tab4} and Fig.\,\ref{fig3}). These emitting regions extend from a region relatively close to the surface of the star (7.5 and 1.5\,R$_*$ for $kT$ of 0.10 keV and 0.20 keV) to very far out in the wind (85 and 38\,R$_*$, respectively). The two hotter plasma components ($kT$ of 0.40 keV and 0.69 keV) start at 2.7 and 3.1\,R$_*$ and extend over a smaller range of radii, out to 4.0 and 4.1\,R$_*$, respectively.

With these results, it appears that the region between 3.0 and 4.0\,R$_*$ contains X-ray emitting plasma with 3 different temperatures.
A priori, in an outwards accelerating wind, one expects the velocity jumps due to instabilities to have a lower amplitude in the inner parts of the wind. Therefore, one expects that the energy available in the shocks is only sufficient to heat the shocked matter to rather low temperatures. Farther out in the wind, the difference of velocity between two consecutive clumps of matter increases and the energy available in the shocks increases accordingly. In the outermost parts of the wind, the density and hence the emission measures become low and the same applies to the velocity jumps since the efficiency of radiative acceleration progressively drops to zero.

From this simple reasoning, one would thus expect the hottest plasma components to arise from intermediate radii in the wind, whilst the softer emission would be produced over a wider ranger of radii starting closer to the stellar surface and extending farther away from it. 

\begin{figure}
\begin{center}
\includegraphics[width=8cm]{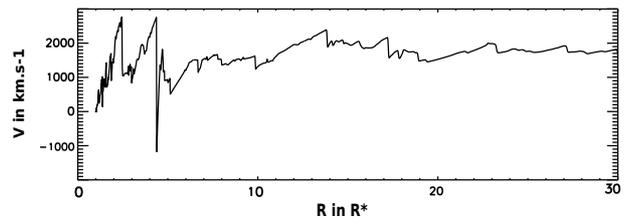}
\caption{Snapshot of the radial velocity in the stellar wind as obtained in 1-D Line Driven Instabilities calculations (Feldmeier 2012, private communication).}
\label{fig5}
\end{center}
\end{figure}

 With the caveat that we currently ignore where the 2.18\,keV plasma, responsible for the high-energy extension of the EPIC and HETG spectra, is located, our results are in qualitative agreement with these expectations as well as with the 1-D line driven instability simulation of Feldmeier et al.\ (\cite{feld97}). These authors show indeed that shocks can arise very close to the stellar surface (see Fig.\,\ref{fig5}, A. Feldmeier, private communication). In our best fit model, the temperature of the hottest plasma component that is present at a given radial position in the wind, increases outwards from the innermost boundary of the emission region (1.5\,$R_*$ where $kT$ reaches 0.20\,keV) to a value of $kT$ = 0.69\,keV in a region between 3 and 4\,R$_*$. Further out, the temperature of the hottest plasma component decreases again. The maximum shock temperature hence occurs in the 3 --  4\,R$_*$ range where the velocity jumps predicted by the simulations reach indeed their highest values. At positions closer to the stellar surface or farther out in the wind, the simulations predict velocity differences between two consecutive clumps that are smaller, in agreement with the lower temperatures found in these regions.

The low value of the filling factor of the $kT$ = 0.69\,keV component indicates that only a fraction of the shocks in the region between 3 and 4\,R$_*$ actually heat the plasma to high temperatures. To ensure that most of the kinetic energy of a fast moving clump which hits a slower clump, moving ahead of it, is indeed converted into heat, the collision must occur along the radial direction. From a statistical point of view, in a 3-D wind, the probability that the velocity vectors of two colliding clumps are not collinear is most probably larger than the probability that they are collinear. Thus only a fraction of the collisions will produce the required heating to reach the hottest temperature.

The overlap between the highest temperature and the lower temperature is somewhat inherent to our methodology. Indeed, whilst we worked with four different temperatures for the X-ray emitting plasma, we consider that the emission region of each plasma component is continuous (i.e.\ there are no gaps) between its inner and outer boundary. For instance, our results indicate the presence of a $kT$=0.10\,keV component relatively close to the star to fit the intensity of the lines at wavelengths above 20\,\AA, but also in the outer parts of the wind to fit the broad wings of the line profiles. In our code we can only assume a continuous shell extending all the way from 7.5 to 85\,R$_*$. The introduction of a discontinuous region would considerably increase the number of free parameters in the regression routine. However, the scenario of an extended, continuous shell for the coolest plasma is supported by two considerations. First, as stated above, non-collinear collisions produce cooler plasma, throughout the whole wind. Second, in a wind embedded plasma, the hottest plasma cools down progressively as it moves outwards with the wind. For instance, adopting the parametrization of the cooling function used by Feldmeier et al.\ (\cite{feld97}), we estimate a typical cooling time of $\sim 180$\,s for the 0.40\,keV plasma at the inner boundary of that component. In the same way, we estimate a cooling time of $\sim 600$\,s for the 0.69\,keV component. On the other hand, the flow times required for the hot gas to move from the inner boundary of its production site (near 2.7 and 3.1\,$R_*$ for the 0.40 and 0.69\,keV plasma respectively) to a location at 5\,$R_*$ are estimated at about 16500\,s and 13500\,s, respectively for the 0.40 and 0.69\,keV gas. Hence the X-ray photons emitted at energies of $kT$=0.10\,keV and 0.20\,keV beyond about 5\, R$_{*}$ exist through a combination of direct emission from weaker shocks, and the cooling of the hot plasma initially heated to 0.40 and 0.69\,keV.

\subsection{Abundances}
Our best-fit model of the RGS spectrum of $\zeta$\,Pup yields an over-abundance of nitrogen along with a depletion of oxygen and carbon (see Table\,\ref{tab2}). Our CNO abundances agree qualitatively with those of the UV/optical studies of Bouret et al.\ (\cite{bouret12}), but are at odds with the result of Pauldrach et al.\ (\cite{pauldrach12}), at least as far as the carbon abundance is concerned. The Ly$\alpha$ doublet of C\,{\sc vi} at 33.73\,\AA\ is the only isolated feature of carbon in the RGS wavelength domain. Nevertheless, it appears impossible to reproduce this line with an over-abundance of carbon as advocated by Pauldrach et al.\ (\cite{pauldrach12}).
 
The other elements, that are relevant for the X-ray spectrum, are assumed to have solar abundances (Andres \& Grevesse,\ \cite{ag89}). We cannot determine the abundances of hydrogen and helium directly with the RGS spectrum since there are no lines of these elements in the X-ray domain. However, they play a key role in the determination of the mass absorption coefficient. Therefore, by default, we use the results of Bouret et al.\ (\cite{bouret12}). 

The particular properties of $\zeta$\,Pup can explain the strongly non-solar CNO abundances. Indeed, $\zeta$\,Pup is a rapidly rotating runaway star and its kinematic properties and chemical abundances could indicate that the star was part of a binary system that underwent either an episode of mass transfer and supernova explosion of the companion or a merger event (Vanbeveren \cite{van11}).

\begin{table*}
\caption{Stellar wind parameters}
\begin{center}
\begin{tabular}{ccccccccc}
\hline
&T  & $v_{\infty}$ & $\beta$ & $R_{*}$  & $\dot{M}$ & X(C) &  X(N) & X(O) \\
&kK & km\,s$^{-1}$ & & $R_{\sun}$ & $10^{-6}\,M_{\odot}$\,yr$^{-1}$ & & & \\
\hline
This work&40$^1$ &  2300 & 1 & 17.3 & 3.5 & $ 6.00 \times 10^{-4}$ & $7.7 \times 10^{-3}$ & $3.05 \times 10^{-3}$ \\
\hline
Bouret et al. 2012 & 40 & 2300 & 0.9 & 17.3 & 2.00  & $2.86 \times 10^{-4}$ & $1.05 \times 10^{-2}$&  $1.30 \times 10^{-3}$\\ 
\hline
Pauldrach et al 2012 &39  &  2100    &  -  &  19    &  13.7 & $8.2 \times 10^{-3}$ & $9.13 \times 10^{-3}$ &  $1.14 \times 10{-3}$   \\
\hline
\end{tabular}
\end{center}
\label{tab2}
\tablefoot{$^1$ We can not determine the effective temperature of the star with our code. As this parameter is important for the mass absorption coefficient calculation, we indicate the value we have adopted in our analysis.
The CNO abundances are given as mass fractions. Note that Pauldrach et al.\ (\cite{pauldrach12}) do not assume a $\beta$-law for the wind velocity, but perform a full hydrodynamical computation of the wind.}
\end{table*}

\subsection{Mass-loss rate and fragmentation}
We find a slightly higher mass-loss rate ($(3.5 \pm 0.25) \times 10^{-6}\,M_{\odot}$\,yr$^{-1}  $) than the UV/optical and IR studies ($(2 \pm 0.2) \times 10^{-6} $ and $2.1 \times 10^{-6}$\,$M_{\odot}$\,yr$^{-1}$ respectively, Bouret et al.\ \cite{bouret12}, Najarro et al.\ \cite{najarro11}) but the difference is not important. Moreover, Bouret et al.\ (\cite{bouret12}) indicate that their mass-loss rate is probably too low to be consistent with the line driving of the wind. Our determination of the mass-loss rate is also in reasonable agreement with the value of $2.5 \times 10^{-6}$\,$M_{\odot}$\,yr$^{-1}$ obtained independently by Lamers \& Leitherer (\cite{LL}) from the radio fluxes of $\zeta$\,Pup and by Oskinova et al.\ (\cite{Oskinova07}) from a fit of the H$\alpha$ line and the P\,{\sc v} resonance doublet.

 Our results indicate that the wind structure of $\zeta$\,Pup is unlikely to be very porous. Indeed, we have tested various assumptions on the porosity parameter: $h' = 0$ (no porosity), $h' = 0.07$ (roughly equivalent to the fragmentation frequency $n_0 = 1.7 \times 10^{-4}$\,s$^{-1}$ derived by Oskinova et al.\ (\cite{OFH}) from a line-by-line investigation of the HETG spectrum) and an intermediate value of $h' = 0.02$ (model$_2$ in Table\,\ref{tab4}). By far the best fit quality is achieved for the model that has no porosity (see Table \ref{tab4}). A similar conclusion was reached by Cohen et al.\ (\cite{cohen10}) who found that porosity was not required to fit individual lines of the HETG spectrum.

From the stellar and wind parameters assumed ($R_*$, $v_{\infty}$, $\beta$) or derived ($\dot{M}$) in this paper, we can estimate what would be the threshold value of the porosity parameter $h'$ for clumps to become optically thick. From Fig.\,\ref{fig1}, we find that $\kappa \sim 170$\,cm$^2$\,g$^{-1}$ at a wavelength of 20\,\AA\ and at 4\,$R_*$. With the volume filling factor of the cool gas ($f = 0.05$) derived by Bouret et al.\ (\cite{bouret12}), we find that a clump size of $l = 0.05\,R_*$, i.e.\ $h' = 0.25$  would be needed to achieve $\tau_c = 1$ at 20\,\AA\ (and at 4\,$R_*$). A value of $h' < 0.02$, as suggested by our fits, implies that $\tau_c = 1$ at 20\,\AA\ would only be achieved in the outermost regions of the wind, where the porosity has increased significantly (assuming, as we have done here, $h(r) = h' \times r$). Therefore, in view of our results, it seems that porosity does not play a role in the formation of the X-ray spectrum of $\zeta$\,Pup. This situation contrasts with the results of Oskinova et al.\ (\cite{WR6}) for the WN5 Wolf-Rayet star WR\,6. The latter authors concluded that the wind of the WN5 star had to have a very porous structure to allow the observed X-ray emission to escape.

\subsection{The {\it{fir}} triplet formation}
In previous studies (Waldron et al.\ \cite{waldron01}, Leutenegger et al.\ \cite{Leutenegger06}, \cite{Leutenegger07}, Oskinova et al.\ \cite{OFH}), the analysis of the {\it{fir}} triplets of the different He-like ions was done line by line in order to determine the hot plasma temperature and the dilution of the UV radiation field in the X-ray emission zone. For a single temperature plasma, this analysis constrains, in principle, the position of the X-ray emitting plasma in the wind.\\

In our analysis, we clearly show that most spectral lines, including the He-like {\it{fir}} triplets, are the result of the combination of emission from several components at different temperatures (Fig.\ref{fig3}). This implies that the analysis of the {\it{fir}} triplets, assuming they are formed by a single temperature plasma and not taking into account the abundances of the elements, could be biased. Let us illustrate this point by the O\,{\sc vii} and N\,{\sc vi} triplets. In Fig.\,\ref{fig4}, one can see that the line profiles and intensities of these triplets are reasonably well fitted. However, Fig.\,\ref{fig3} shows that both triplets are actually formed by contributions of the two lowest temperature plasmas which produce quite different line profiles and intensities. In the case of the N\,{\sc vi} triplet, for instance, the inter-combination line is stronger than the resonance line for the 0.10\,keV plasma component (in red in Fig.\ref{fig3}), whilst the reverse situation applies to the 0.20\,keV component. The combination of both components fits the observed triplet quite well. 

\begin{figure}
\begin{center}
\includegraphics[width=8cm]{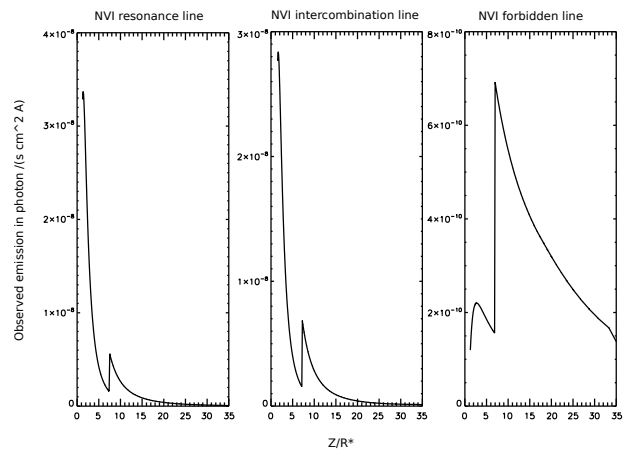}
\caption{Contribution to the observable emission for the {\it fir} components of the N\,{\sc vi} He-like triplet. The product of the intrinsic emission per solid angle with $exp{(-\tau)}$ is shown as a function of $z$ (in $(p,z)$ coordinates with $p = 0.5$\,R$_*$). The triplet includes contributions from the 0.10 and 0.20\,keV plasma components starting respectively at $r = \sqrt{p^2 + z^2} = 7.5$\,R$_*$ and $r = 1.5$\,R$_*$.}
\label{figN6}
\end{center}
\end{figure}

\begin{figure}
\begin{center}
\includegraphics[width=8.5cm]{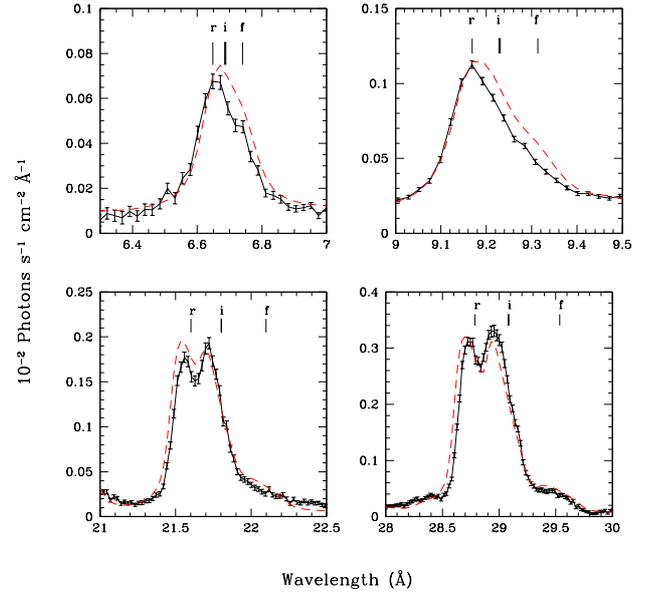}
\caption{ Our best-fit (dashed red line) of four {\it fir} triplets present in the RGS spectrum of $\zeta$\,Pup (in solid black line).}
\label{fig4}
\end{center}
\end{figure}

Figure\,\ref{figN6} illustrates the contributions to the observed emission of each component of the N\,{\sc vi} triplet for a line of sight of impact parameter $p = 0.5$\,R$_*$. Owing to the pumping by the UV photons from the upper level of the {\it f} transition, the latter line is strongly surpressed at the benefit of the {\it i} line. This is especially true in the innermost regions of the wind, where the 0.20\,keV plasma component is the sole contributor to the N\,{\sc vi} triplet. As a result, ${\cal G} = \frac{f + i}{r} \simeq \frac{i}{r}$ and the ratio between the strength of the inter-combination line and the resonance line tends towards the value of ${\cal G}$ predicted by the atomic data.

We further note the presence of numerous weak lines in the AtomDB emissivities. Due to the Doppler broadening of the lines and to the limited spectral resolution of the RGS instrument, these weak lines cannot be seen individually but they contribute to the total emission. This is particularly important in the inter-combination line of N\,{\sc vi}. 

As can be seen on Figs.\,\ref{fig2} and \ref{fig4}, our model has some of its strongest deficiencies in the {\it fir} triplets. For instance, the model predicts too large a flux in the red wind of the blends of the Mg\,{\sc xi} and Si\,{\sc xiii} triplets (Fig.\,\ref{fig4}), which likely indicates an overestimate of the inter-combination (and possibly forbidden) line with respect to the resonance line. Likewise, the resonance lines of the N\,{\sc vi} and O\,{\sc vii} ions are predicted with slightly too strong a blue-shift compared to the observations. There are several possible reasons for these discrepancies.

Clearly, the sampling of the emitting plasma into four discrete temperature components is a simplification that affects the results. The actual plasma in the wind of $\zeta$\,Pup is more likely to have a roughly continuous temperature distribution. 

Another point is the radial dependence of the hot plasma filling factor. In our model, we assume that the filling factor of a given plasma component remains constant over the full extent of the emission region of this component. This might be an over-simplification, as the filling factor could change, especially for the 0.10 and 0.20\,keV plasma components that span a wide range of radii. Although implementing a radial dependence of the filling factor into our code is in principle possible, we have refrained from doing so, as it would introduce additional free parameters, thereby rendering the search for a best fit model even more difficult. 

A third avenue to explore could be the velocity-law of the hot gas. Throughout this paper, we have assumed that the X-ray emitting plasma follows the same $\beta = 1$ velocity-law as the cool gas. However, the hot plasma is highly ionized, thus lacking line opacity for the radiative acceleration by the stellar UV radiation. Although Coulomb forces help to carry the hot gas along with the cool wind, one cannot exclude the possibility that the velocity law might differ from the one of the cool wind. Testing the impact of the velocity law on our fits is beyond the scope of the present paper. 

Finally, more severe discrepancies, of similar type as those observed here, motivated Leutenegger et al.\ (\cite{Leutenegger07}) to incorporate the effect of resonant scattering into their line-by-line fits of the N\,{\sc vi} and O\,{\sc vii} triplets\footnote{Leutenegger et al.\ (\cite{Leutenegger07}) further noted that models without resonant scattering produce an inter-combination line that was not sufficiently blue-shifted compared to the observations (their Figs.\,1 and 2). The latter discrepancy is less severe in our model thanks to the combination of several plasma components with different temperatures and locations.}. Using the characteristic Sobolev optical depth $\tau_{0,*}$ as a free parameter, Leutenegger et al.\ (\cite{Leutenegger07}) were indeed able to improve the quality, of their fits of these two triplets, but at the expense of very large, sometimes infinite, values of $\tau_{0,*}$. Such large characteristic Sobolev optical depths are problematic though, as they require a filling factor of the hot plasma near unity, far away from the values derived here and by other studies (Hillier et al.\ \cite{Hillier93}). This problem was noted by Leutenegger et al.\ (\cite{Leutenegger07}), who advocated that such high filling factors were only required in small regions where line formation takes place. However, such a configuration would then also have a strong impact on the formation of other lines, which is not seen. This is why we decided not to include resonance scattering into our model and do not consider this effect as a good candidate for solving the observed discrepancies.

\begin{figure*}[h!tb]
\begin{center}
\includegraphics[width=15cm]{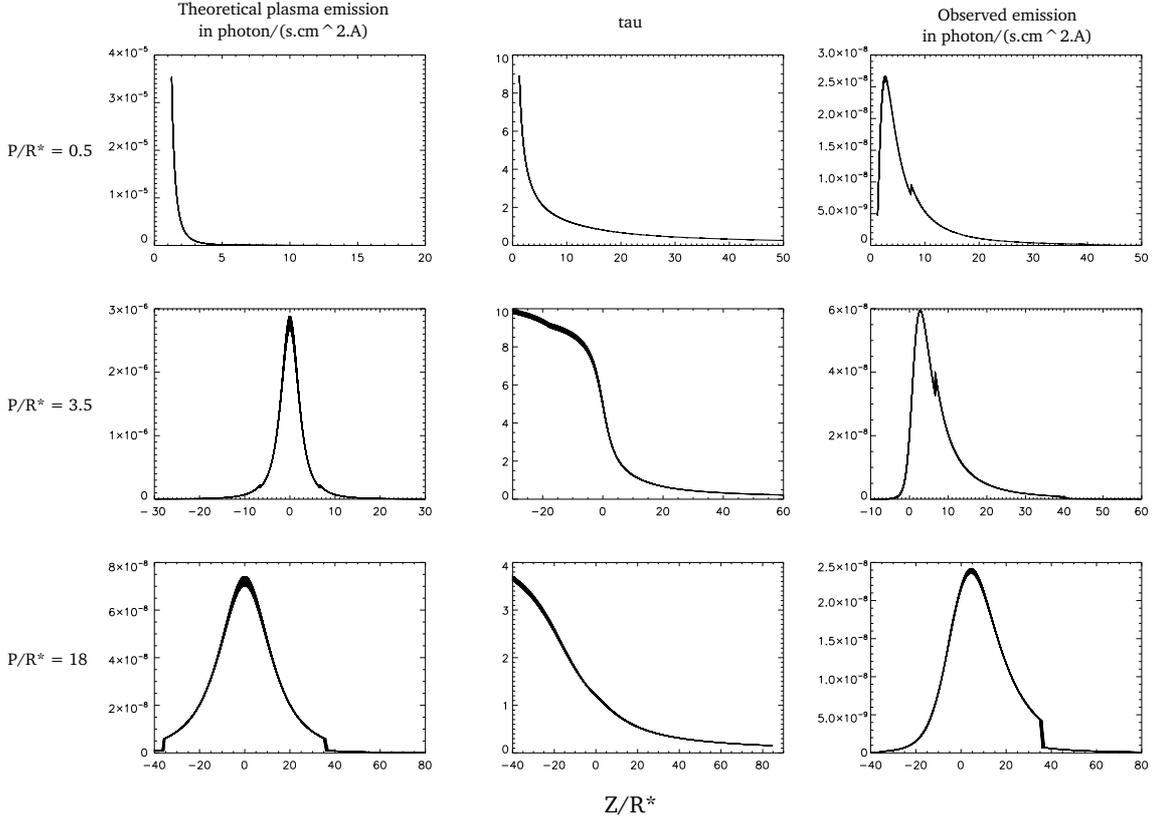}
\caption{ Line formation region of the O\,{\sc vii} resonance line. The left panels yield the variation of the intrinsic emission per solid angle (emissivity multiplied by the density squared) along the $z$ axis of the $(p, z)$ coordinates for different values of the impact parameter $p$. The middle panels provide the optical depth $\tau$ along the line of sight to the observer (located at $z \rightarrow \infty$). The right panels yield the contribution to the observable emission, i.e.\ the product of the intrinsic emission per solid angle with $\exp(-\tau)$. The morphology of this distribution results from the combination of the two plasma components at 0.10 and 0.20\,keV that extend between 7.5 and 85\,$R_*$ for the former and 1.5 and 38\,$R_*$ for the latter.}
\label{lineformation}
\end{center}
\end{figure*}

Using the ${\cal R}$ ratios evaluated on the HETG spectra of $\zeta$\,Pup, Cassinelli et al.\ (\cite{cassinelli01}), estimated a radial range between 1.9 and 4.0\,$R_*$ and between 1.7 and 2.25\,$R_*$ for the formation regions of the Mg\,{\sc xi} and Si\,{\sc xiii} triplets respectively. Based on the same dataset, Leutenegger et al.\ (\cite{Leutenegger06}) found an inner emitting radius of the Mg\,{\sc xi} and Si\,{\sc xiii} triplets of $(1.43 \pm 0.10)\,R_*$, whilst Oskinova et al.\ (\cite{OFH}) inferred an inner radius of $\simeq 1.5\,R_*$. At first sight, these results of Cassinelli et al.\ (\cite{cassinelli01}) for the Mg\,{\sc xi} triplet are in reasonable agreement with the location of the 0.40\,keV plasma component, which dominates the formation of these lines in our model (see Fig.\,\ref{fig3}). This good agreement between the local analysis of Cassinelli et al.\ (\cite{cassinelli01}) is due to the narrow radial domain covered by the 0.40\,keV component. In contrast, Cassinelli et al.\ (\cite{cassinelli01}) found that the Ne\,{\sc ix} and O\,{\sc vii} triplets were likely formed below 4 and 10 stellar radii respectively. At first sight, this result disagrees with the location of the 0.20\,keV plasma component which extends from 1.5 to 38\,$R_*$. However, the actual location of the formation region of the observable lines results from the interplay between intrinsic emission and absorption, which is rather complex (see Fig.\,\ref{lineformation}). For instance, the O\,{\sc vii} resonance line has its observable emission arising mainly from below 15\,$R_*$ in not too bad agreement with the results of Cassinelli et al.\ (\cite{cassinelli01}).

For the S\,{\sc xv} triplet around 5.04 - 5.10\,\AA, Leutenegger et al.\ (\cite{Leutenegger06}) obtained an inner radius of only $(1.1^{+0.4}_{-0.1})\,R_*$, i.e.\ potentially much closer to the photosphere than any of the emitting regions in our model. However, the HETG spectrum analysed by these authors is of rather poor quality at these wavelengths. Unfortunately, the RGS instrument does not cover this wavelength domain and we cannot verify whether or not the morphology of this line is well predicted by our model. 

\subsection{The Fe\,{\sc xvii} $\lambda$\,17.096\,\AA\ line}
\begin{figure}
\begin{center}

\includegraphics[width=7cm]{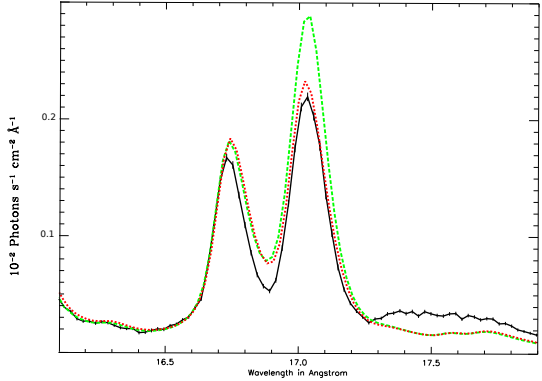}
\caption{Modification of the theoretical line profile and intensity of the Fe\,{\sc xvii} $\lambda$\,17.096\,\AA\ line due to the UV radiative flux. The observations are in solid black line, the theoretical prediction without UV pumping is in dashed green line and our best-fit model with an ad hoc reduction of the line due to UV pumping is in dotted red line.}
\label{fig6}
\end{center}

\end{figure}

The forbidden lines of the He-like ions are not the only spectral features arising from metastable levels which are depopulated by the strong UV radiation field. Indeed, the transition Fe\,{\sc xvii} $\lambda$\,17.096\,\AA\  also arises from a metastable level. Mauche et al.\ (\cite{mauche01}) have shown that this level can be depopulated by a strong extreme-UV continuum radiation between 190 and 410\,\AA. They also found that in the case of a stellar surface temperature $\sim$ 55\,kK, the level could be totally depopulated. Since our code does not include a full treatment of the atomic transitions of the Fe\,{\sc xvii} ion and since we have no observational constraints on the extreme UV emission of $\zeta$\,Pup, we have included an adjustable reduction factor to fit the intensity of this line. Our best-fit results correspond to a division of the nominal line intensity by a factor 4.5 (see Fig.\,\ref{fig6}). Therefore, it seems that the metastable level is heavily depopulated, which is not surprising as the effective temperature of $\zeta$\,Pup is 40\,kK. The emission seen in Fig.\,\ref{fig6} close to 17.1\,\AA\ actually corresponds to the Fe\,{\sc xvii} $\lambda$\,17.051\,\AA\ line.

\subsection{Additional unsolved problems in the fit}
Beside the discrepancies in the fits of the {\it fir} triplets discussed above, a few additional regions in the spectra are not well fitted. The main one is the region between 17.2 and 17.65\,\AA, where the flux is under-estimated in our model (see Fig.\,\ref{fig6}). The atomic database of the AtomDB model does not contain any transitions which could easily explain this strong emission in the observation. We have also checked whether there are any radiative recombination continua in that spectral range, but we could not find any. An instrumental effect can be ruled out as a similar bump is seen in the {\it Chandra} HETG spectrum of $\zeta$\,Pup (Cassinelli et al.\ \cite{cassinelli01}, their Fig.\,1) and Waldron et al.\ (\cite{waldron01}, see Fig.2 in their paper) also find a similar structure in the {\it Chandra} HETG spectrum of $\zeta$\,Ori.

The problem could be due to an underestimate of the O\,{\sc vii} series that affects this part of the spectrum. Although the intensity of the {\it fir} triplet is fitted correctly, an underestimate of the strength of the O\,{\sc vii} edge near 20.8\,\AA\  in our model could actually lead to an underestimate of this series. Alternatively, the problem could point at a lack of atomic data of some elements or some peculiar ionization stages which are complicated to calculate. The failure to reproduce the flux level in this region has also certainly an impact on the determination of the attenuation factor of the Fe\,{\sc xvii} $\lambda$\,17.096\,\AA\ line as well as on the under-estimated continuum beetween 18 and 19\,\AA. We do not think that this bump could result from a redistribution of the flux of the Fe\,{\sc xvii} $\lambda$\,17.096\,\AA\ transition as no such feature is seen in other spectra where pumping is very effective (Mauche et al.\ \cite{mauche01}).

We encounter a similar problem in the region between 12.3 and 13\,\AA\ and for two more 'lines' at $\sim$ 10.0 and 11.0\,\AA. If there exists another plasma component, hotter than those included in our model, in the wind of $\zeta$\,Pup, some of these features could actually be due to numerous weak lines of higher ionization species of iron. However, with our current four-component model, these lines are not predicted.

\section{Summary and conclusions}
In this paper, we have presented a multi-temperature embedded plasma model that fits the RGS spectrum of $\zeta$\,Pup. Figure\,\ref{fig2} shows that the intensity of the strongest lines is well reproduced, although there remain some minor differences. As for any spectral synthesis code, our method relies on some assumptions that introduce some limitations. For instance, one of these limitations is the discretization of the temperature structure of the hot plasma in the wind, as we use `only' four temperatures. Moreover, our temperature grid itself is also discretized with a step in $kT$ of 0.025\,keV. As the line intensities and the ${\cal R}_0$ ratio for the treatment of the {\it fir} triplets are dependent on the temperature, the combination of our assumptions in the entire fitting procedure can probably explain at least some of the remaining small differences between the observed RGS spectrum of $\zeta$\,Pup and our best-fit model. Additional aspects to be explored in the future are the radial dependence of the hot plasma filling factor and the velocity law of the hot gas.
We further assumed a simplified radial dependence of the mass absorption coefficient and used the bridging law of Owocki \& Cohen (\cite{OC06}) to calculate the optical depth from the optically thick to optically thin regime.
Finally, we stress that, even the best-quality atomic data have typical uncertainties of 20\%. Some of the differences between our model and the observations could thus result from the uncertainties on the atomic parameters.

Nevertheless, our code provides a major step forward in the analysis of the X-ray spectra of O-type stars. Indeed, we are now in the position to fit the full X-ray spectrum consistently. From these fits, we derive the temperature distribution in the wind, and constrain the mass-loss rate as well as the abundances of a number of elements.

To the best of our knowledge, our results provide the first empirical determination of the temperature distribution of the X-ray emitting plasma in the wind of a massive star. In our analysis, the shocks producing X-ray emission arise close to the surface of the star (1.5\,R$_*$) and extend over a large zone (up to about 85\,R$_*$). However the highest temperature ($kT$= 0.69 keV) is produced in a smaller region, at 3 -- 4\,$R_*$, in agreement with 1D hydrodynamical calculations.

Another important result of our work is our finding that porosity does not improve the quality of the fit of the global spectrum. It seems thus that the clumps that make up the wind of $\zeta$\,Pup must have rather small dimensions.

Concerning the He-like {\it fir} triplets, we have shown that analyses that neglect the multi-temperature nature of the emitting plasma could be biased when attempting to determine the inner radius of the emission region of the plasma. This is due to the fact that several temperatures contribute to the line emissions.

We find a depletion of carbon and oxygen and an over-abundance of nitrogen. We also determine a mass-loss rate of $3.5 \times 10^{-6}\,M_{\odot}$\,yr$^{-1}$ for $\zeta$\,Pup. These results are in agreement with recent determinations by Bouret et al.\ (\cite{bouret12}) and Najarro et al.\ (\cite{najarro11}), but disagree with the analysis of Pauldrach et al.\ (\cite{pauldrach12}).

In future studies, we will apply our code to other O-type stars with good quality high-resolution X-ray spectra to generalize the results obtained for $\zeta$\,Pup in the present paper.

\acknowledgement{This research is supported by the Communaut\'e Fran\c caise de Belgique - Action de recherche concert\'ee (ARC) - Acad\'emie Wallonie--Europe and by an XMM+INTEGRAL PRODEX contract (Belspo). YN acknowledges support by the FNRS (Belgium). Many thanks to John Hillier for making his code CMFGEN available to the community and to A. Feldmeier for providing some results of his LDI simulations.}

\end{document}